\documentclass[12pt]{article} \usepackage{latexsym}
\usepackage{amsfonts} \usepackage{amsmath} \oddsidemargin 0in
\begin{document}


\newcommand{\olc}{\ensuremath{\mbox{\Large $\prec $\normalsize}}}
\newcommand{\orc}{\ensuremath{\mbox{\Large$ \succ$\normalsize}}}
\newcommand{\lc}[1]{\ensuremath{{}^{{}^{(#1)}}\!\!\!\!\!\olc}}
\newcommand{\rc}[1]{\ensuremath{\orc\!\!\!\!\!^{{}^{(#1)}}}}
\newcommand{\Ax}{\ensuremath{A^{\times}}}
\newcommand{\Phix}{\ensuremath{\Phi^{\times}}}
\newcommand{\APhi}{\ensuremath{A^{\dag}|_{\Phi}}}
\newcommand{\ip}[2]{\ensuremath{(#1,#2)}}
\newcommand{\bk}[2]{\ensuremath{\langle #1|#2 \rangle}}
\newcommand{\norm}[1]{\ensuremath{\left|#1\right|^{2}}}
\newcommand{\kt}[1]{\ensuremath{|#1\rangle}}
\newcommand{\br}[1]{\ensuremath{\langle #1|}}
\newcommand{\bohmHS}{\ensuremath{{\cal H}}}
\newcommand{\kb}[2]{\ensuremath{| #1\rangle\langle #2|}}
\newcommand{\ktbr}[2]{\ensuremath{| #1\rangle\langle #2|}}
\newcommand{\vpe}{\ensuremath{\phi(E)}}
\newcommand{\gktp}{\ensuremath{|E_{R} - i\Gamma/2\rangle}}
\newcommand{\psig}{\ensuremath{\psi_{G}}}
\newcommand{\mom}{\ensuremath{\mathbf{p}}}
\newcommand{\mate}[3]{\ensuremath{\langle #1|#2|#3 \rangle}}
\newcommand{\dbk}[2]{\ensuremath{( #1|#2 )}}
\newcommand{\dkt}[1]{\ensuremath{|#1)}}
\newcommand{\dbr}[1]{\ensuremath{( #1|}}
\newcommand{\dkb}[2]{\ensuremath{|#1)(#2|}}
\newcommand{\pbk}[2]{\ensuremath{\langle {}^{+}#1|#2{}^{+} \rangle}}
\newcommand{\pkt}[1]{\ensuremath{|#1{}^{+}\rangle}}
\newcommand{\pbr}[1]{\ensuremath{\langle {}^{+}#1|}}
\newcommand{\pkb}[2]{\ensuremath{| #1{}^{+}\rangle\langle {}^{+}#2|}}
\newcommand{\mbk}[2]{\ensuremath{\langle {}^{-}#1|#2^{-} \rangle}}
\newcommand{\mkt}[1]{\ensuremath{|#1^{-}\rangle}}
\newcommand{\mbr}[1]{\ensuremath{\langle {}^{-}#1|}}
\newcommand{\mkb}[2]{\ensuremath{| #1{}^{-}\rangle\langle {}^{-}#2|}}
\newcommand{\mpbk}[2]{\ensuremath{\langle {}^{-}#1|#2^{+} \rangle}}
\newcommand{\pmbk}[2]{\ensuremath{\langle {}^{+}#1|#2^{-} \rangle}}
\newcommand{\mpkb}[2]{\ensuremath{| #1{}^{-}\rangle\langle {}^{+}#2|}}
\newcommand{\pup}[1]{\ensuremath{#1^{+}}}%
\newcommand{\mup}[1]{\ensuremath{#1^{-}}}%
\newcommand{\pdo}[1]{\ensuremath{#1_{+}}}%
\newcommand{\mdo}[1]{\ensuremath{#1_{-}}}%
\newcommand{\tup}[1]{\ensuremath{#1^{\times}}}%
\newcommand{\pmdo}[1]{\ensuremath{#1_{\pm}}}
\newcounter{vnstuff}
\newcounter{vns}
\newcounter{gv}

\begin{titlepage}
\begin{center}

\huge \textbf{Quantum Theory in the Rigged Hilbert
Space---Irreversibility from Causality
\footnote{to appear in \emph{Irreversibility and Causality
in Quatum Theory: Semigroups and Rigged Hilbert Space
}, Arno Bohm, H.D. Doebner, P. Kielanowski, eds; Vol. 504
Springer Lecture Notes in Physics}} \vspace{4pt}\\ \Large
\textbf{A.~Bohm and N.~L.~Harshman} \vspace{4pt}\\ \small{e-mail:
bohm@physics.utexas.edu, harshman@physics.utexas.edu}\\ \small{Physics
Department}\\ \small{The University of Texas at Austin}\\
\small{Austin, Texas~78712}

\begin{abstract}
After a review of the arrows of time, we describe the possibilities of 
a time-asymmetry in quantum theory.  Whereas Hilbert space quantum
mechanics is time-symmetric, the rigged Hilbert space formulation, which arose 
from Dirac's bra-ket formalism, allows the choice of asymmetric boundary 
conditions analogous to the retarded solutions of the Maxwell equations for
the radiation arrow of time.  This led to irreversibility on the 
microphysical level as exemplified by decaying states or resonances.
Resonances are mathematically represented by Gamow kets, 
functionals over a space of very well-behaved (Hardy class) vectors, which
have been chosen by a boundary condition (outgoing for decaying states).  
Gamow states have all the properties that one heuristically needs for 
quasistable states.  For them a Golden Rule can be derived from the 
fundamental probabilities ${\cal P}(t)=\mathrm{Tr}(\Lambda(t)
W^{\mathrm{Gamow}}(t_0))$ 
that fulfills the time-asymmetry condition $t\geq t_0$ 
which could not be realized in the Hilbert space.
\end{abstract}
\end{center}
\end{titlepage}

\section{Preface---Time-Asymmetries}

This preface was added to the 
paper in order to explain what we mean by the word irreversible in the title.

The term ``irreversible'' has two different uses and has been applied to
several 
different phenomena~\cite{pheno,Davies}.  These different phenomena are also 
called different
 ``arrows of time''.  These arrows are not unrelated to each other,
but there is no consensus as to their exact relationships.  Somehow, all
these arrows seem to be connected to the vague, intuitive notion of 
causality.

Most of the 
time the word ``irreversibility'' is used to refer
 to these arrows of time, i.e. to
the directedness of the time evolution
of a physical system or of the state of a physical system, classical or
quantum. An alternate use of the word ``irreversible'' is 
to describe non-invariance (of the observables) 
with respect to the (antilinear) time reversal transformation $A_T$.
These two notions, though related 
are not the same and 
must be distinguished.  Whereas the time reversal operator is defined by its
 relations to 
the other {\em observables} that generate the space-time symmetry 
transformations or by its relations with these transformations (Poincar\'e 
group,
 Galilean group), ``irreversibility'' above means the impossibility or 
improbability to create
 a {\em state} which evolves backward in time.  
Irreversible time evolution is not necessarily in contradiction with a 
$T$-invariant Hamiltonian if a mathematical theory is used 
that makes a distinction between states and observables.  
Time reversal invariance---or its violation---is
not the key for understanding irreversibility.  We shall mention
 $A_T$ transformations only briefly in Sect.\ \ref{Semigr}.  In this paper and 
in the title, irreversibility is the time-asymmetry due to a preferred
direction of time evolution.

Of the different irreversible phenomena or the different ``arrows of time'',
most prominent and oldest is the thermodynamic 
arrow of
time (TA): the 
entropy ${\cal S}$ in an isolated (classical) physical system increases,
$\frac{d{\cal S}}{dt}>0$, until it reaches equilibrium, 
$\frac{d{\cal S}}{dt}=0$ (2nd
law of thermodynamics).  Using (for the sake of simplicity) the time-symmetric
equation of Newtonian mechanics for the scattering of molecules in a gas,
it has been shown~\cite{quote} that Boltzmann's Stosszahl-Ansatz implies
assumptions (boundary conditions) from which the TA follows: ``If
 one assumes that the gas was prepared in some manner in the past$\ldots$,
then it follows that correlations between molecules and scattering centers
will arise only from past but not from future collisions,'' (R.\ Peierls in
\cite{quote}).  That initial conditions, 
not final conditions, are specified is part of the 
intuitive law of causality.

Another arrow of time of classical physics is the
radiation arrow (RA).  The Maxwell equations, like all local physical laws,
are symmetrical in time, yet retarded radiation predominates.  The 
phenomenological law according to which nature favors the retarded potential 
over the advanced solution of the Maxwell field equations is called the
time arrow of radiation.  An effective way to describe this time arrow
is to formulate an additional axiom to the Maxwell equations---a particular,
time-asymmetric, boundary condition.  This boundary condition excludes the 
strictly incoming fields $A^\mu_{in}$ (Sommerfeld radiation condition).  For
 a system of charged particle, the fields 
$A^\mu(x)=A^\mu_{adv} + A^\mu_{out} = A^\mu_{ret} + A^\mu_{in}$ acting on a
particle at $x$, in {\em every} region of space-time, are only the retarded 
fields of the other particles in the region, $A^\mu_{in}=0$.  In other words,
all fields
 possess advanced sources somewhere in the universe;  {\em Radiation 
must first be emitted 
by a source before it can be detected by the receiver}.  Other boundary
conditions are imaginable, e.g.~the time-symmetric boundary conditions for 
which all fields also possess sinks and will be absorbed somewhere.  In the
famous Einstein-Ritz arguments~\cite{EandR},  
Einstein thought that physics could be restricted to the time-symmetric case 
for which retarded and advanced fields are equivalent.  Then the RA 
is based on probability (i.e., the TA). Ritz considered this restriction of the
boundary conditions as not allowed, in which case phenomena demand the choice
of the retarded fields as the only possibility and the TA has its origin in the RA.

The classic mathematical attempt to settle this argument and account for the
 predominance of retarded 
radiation was given by Wheeler and Feynman\cite{FandW}.  They start with
time-symmetric boundary conditions for the field equations (time-symmetric
electrodynamics) and introduce a cosmic absorber (of huge amounts of
dust matter) which annihilates the 
advanced fields over the retarded.  This gives the impression that ``~the
physics of radiation can be regarded as, at bottom, time-symmetric with
only the statistics of large numbers giving the appearance of 
asymmetry~'' (J.A. Wheeler in \cite{pheno}).  However such an absorber has
to contain some arrow of time.  In order to provide the appropriate 
thermodynamic conditions for absorption to occur in the far future, a random 
initial state for the matter has to be postulated, and this initial state
depends in turn on the retardation.  Therewith one is back to the 
Einstein-Ritz controversy~\cite{EandR} as to whether
irreversibility is exclusively based on probability considerations,
$\mbox{TA} \Rightarrow \mbox{RA}$, (which Einstein 
believed) or whether an initial condition and thus causality is the basis
of irreversibility, in which case  
$\mbox{RA} \Rightarrow \mbox{TA}$ (Ritz's opinion).
Up to today, there seems to be no agreement in the literature as to whether
$\mbox{TA} \Rightarrow \mbox{RA}$ or $\mbox{RA} \Rightarrow \mbox{TA}$,
though the former is more prevalent~\cite{pheno}.
Causality and probability may just be two aspects of one and the same
principle.

For every situation
with increasing entropy and retarded potentials one also has a completely
time reversed situation with decreasing entropy and advanced potentials.
This dichotomy is not only restricted to the electromagnetic field but can
also be found in quantum physics.   This is not surprising because the 
quantum theory of radiation cannot be expected to be free of an arrow of 
time if classical electrodynamics 
possesses one.  However, the mathematical theory of the Hilbert space allows 
only time-symmetric boundary conditions.  To accommodate time-asymmetric
boundary conditions one has to extend the mathematical theory beyond the 
confines of the Hilbert space.

In addition to the two arrows of classical physics of relevance to time
scales achievable in the laboratory, a third arrow exists at the cosmic
time scale---the cosmological arrow of time (CA).  This states that the
 universe expands (the contracting solutions of the equation 
of motion are excluded) or at least that we live in the expanding phase
 fairly close 
to the initial singularity of the big bang.

Usually the cosmological arrow is believed to be the master arrow from which 
the others follow [P.C.W. Davis in \cite{pheno}]. 
Though there seems to be no consensus whether 
$\mbox{CA}\Rightarrow\mbox{RA}\Rightarrow\mbox{TA}$ or whether 
$\mbox{CA}\Rightarrow\mbox{TA}\Rightarrow\mbox{RA}$, the most
attractive scenario is the entropy gap: the expansion of the 
universe during the first three minutes was much faster than the
relaxation time of the nucleosynthesis, leaving the majority of the
cosmological material for aeons in a metastable state.  The quasistable 
states of our sun are the metastable stuff that is driving the 
time-asymmetric processes.  On the other hand,
local arrows of time need not depend on asymmetric cosmologies.  For example,
in some simple models~\cite{suss}, the light cones in a closed Friedmann 
universe tend to imitate the expansion figure of space-time.  As a 
consequence, outgoing electromagnetic waves occur near the
big bang and incoming (in our local sense of 
time) waves occur close to the big crunch.

The CA and the big bang give us a means of 
defining the cosmic time and its origin $t=t_0\equiv 0$.  In order to define
irreversible time evolution (time evolution that does not extend to 
$-\infty$ {\em and} to $+\infty$) one has to be able to choose such 
a reference time $t_0\neq \pm \infty$ like the creation time 
of the universe.  In
the laboratory one fixes this time $t_0=0$ for each physical system
individually as, for example, the time at minimum entropy or the time at 
which the radiation is emitted.  In quantum physics (the concern 
of this article) this time $t_0$ is the time at which the
state has been prepared (e.g.~a resonance has been ``created'') 
and at which
one can start the detection of an observable 
(e.g.~of the decay products) in
this state.

Before we turn to the quantum mechanical arrow of time we want to mention the 
psychological arrow.  It is the arrow of time by which we 
remember the past and predict the future.  It is physically not well-defined, 
but usually considered subsumed under the thermodynamic arrow which in turn 
is maintained by the expanding universe.  Thus if in a (re)contracting 
universe the entropy decreases (which may or may not be the case) then the 
psychological arrow should also turn around to be again in
 the direction of entropy increase (Hawking in \cite{pheno}).  
That is, in such a contracting 
universe---or contracting region of the universe---the direction of the
psychological arrow 
should be the opposite of {\em our}
 subjective sense of time; 
initial conditions should turn into final conditions and vice versa.  This 
then 
would be the realm of states with negative time (with $t=0$ meaning
 the creation time of our universe), or the time beyond 
the ``switch-over'' point of maximal expansion, if we think of the cosmic time
as cyclic.  
How this reversal of time could take place 
is hard to comprehend.  It does not seem to be of much practical importance;
probably it is just a feature of the time-symmetric differential equations, 
which are incomplete without boundary conditions.  For the mathematical 
boundary condition we have a choice, but for the physical boundary condition in
our world we have not.  The physics of our universe is time-asymmetric.  It is
important to have a mathematical theory that is not too restrictive and allows
one arrow of time.  Then the opposite arrow of time is obtained by time 
reversal transformation of the boundary condition.  Whether for every physical
system or process there is also a time reversed one in {\em our} world, is a 
different question which is not answered by the T- or CPT-invariance of the
Hamiltonian (differential operator).  Solutions of time-symmetric dynamical 
equations with time-asymmetric boundary conditions come in pairs.
  With the choice of the boundary 
condition, one of the two time-asymmetric solutions is selected.
This applies to the classical equation of general relativity 
(big bang--big crunch; black hole--white hole) and electromagnetism 
(retarded--advanced) and must as well apply to the mathematical theory
of quantum physics.  One should not restrict the 
mathematics of quantum mechanics to such an extent that time-asymmetric 
boundary conditions cannot be formulated with mathematical rigor.  The 
Hilbert space theory of quantum mechanics does not allow such a 
time-asymmetric formulation, while the rigged Hilbert space theory allows
time-asymmetry in either of the directions (which are related by T (or CPT) 
conjugation, cf.~Sect. \ref{Semigr}
and \cite{Bohm4}). The physics in our world chooses one of these directions.

In analogy to the radiation arrow of time, the quantum mechanical arrow of time
 can be formulated without reference to the mathematical theory as: 
{\em a state 
first must be prepared by a preparation apparatus before an observable
can be detected in it by the registration apparatus}.  We will call this the
preparation$\rightarrow$registration arrow of time~\cite{Ludwig}.  The 
preparation$\rightarrow$registration arrow of time is perhaps the closest
we can come to a physical definition of the psychological arrow of time, 
because it is the experimentalist who first prepares the state
and then activates the detector.  Irreversibility that we 
mean in the title is the asymmetry of the time evolution based on this 
arrow of time.  Its most prominent realization in quantum physics is the 
intrinsic time evolution of resonances and decaying states.

In contrast, conventional
irreversibility in quantum physics has been considered to be 
extrinsic.  Two forms of state changes are usually discussed in quantum 
mechanics: 
\begin{enumerate}\item A unitary time evolution 
generated by the Hamiltonian $H$ of the quantum system $S$:
\begin{subequations}\label{evol}
\begin{equation}\label{evola}
\rho \rightarrow \rho(t) = e^{-iHt}\rho(0)e^{iHt} = U^{\dag}(t)\rho U(t),
\; -\infty < t < \infty
\end{equation}
or
\begin{equation}\label{evolb}
\phi(t) = e^{-iHt}\phi,\; -\infty < t < \infty
\end{equation}
where $\rho$ is the statistical or density operator of a mixed state and $\phi$
 the state vector if the state is pure, $\rho = |\phi\rangle\langle\phi|$.

This change of the state is reversible and
 could be equivalently described by
 the Schr\"{o}dinger equation
\begin{equation}\label{evolc}
i\frac{d\phi}{dt}=H\phi
\end{equation}
\end{subequations}
{\em with} the ``Hilbert space'' boundary condition $\phi(t)\in{\cal H}$.
\item The ``reduction of the state vector'' on measurement, 
which in von 
Neumann's idealization is given by the following 
additional axiom (``collapse of the 
wave function''---ideal measurement of 1st kind): If $P_{a_i}$ are the 
projection operators of the observable measured 
$A=\sum_i a_i P_{a_i}$, then
\begin{subequations}\label{collax}
\begin{equation}\label{measa}
\rho\;\rightarrow\;\rho^{\mathrm{after}}(t_0)=
\sum_i P_{a_i} \rho(t_0) P_{a_i}
\end{equation}
or, if the results of the measurement is $a_j$,
\begin{equation}\label{measb}
\rho\;\rightarrow\;\rho^{\mathrm{after}}(t_0)=
P_{a_j} \rho(t_0) P_{a_j}.
\end{equation}
\end{subequations}
Here $\rho^{\mathrm{after}}$ is the state after the measurement.  If one 
defines the (von Neumann) entropy by
\begin{equation}\label{entrop}
{\cal S}(\rho)=-\mathrm{Tr}(\rho\ln\rho)
\end{equation}
then for (\ref{measa})---but not for (\ref{measb})---one obtains:
 ${\cal S}[\rho^{\mathrm{after}}(t_0)]> 
{\cal S}[\rho^{\mathrm{before}}(t_0)]$, 
i.e. this entropy increases if the initial entropy is fixed 
and smaller than the maximal entropy.  If the initial state in (\ref{measa})
 was a pure state $\rho = \ktbr{\phi}{\phi}$, then the state after
measurement is a mixture:
\begin{equation}\label{mixt}
\ktbr{\phi}{\phi} = \sum_i\sum_j |a_i)(a_i\ktbr{\phi}{\phi}a_j)(a_j|
\rightarrow  \rho^{\mathrm{after}} = \sum_i 
|a_i)(a_i\ktbr{\phi}{\phi}a_i)(a_i|.
\end{equation}
This decoherence (elimination of the interference terms 
$(a_i\ktbr{\phi}{\phi}a_j)$, $i\neq j$) 
is considered an irreversible process, though no time
has actually elapsed.
\end{enumerate}

The increase in von Neumann entropy is conventionally 
considered as the 
arrow of time in quantum mechanics.  It is not due to time-asymmetric
laws (equations and/or boundary conditions) but appears through the 
extraneous act of measurement.  The intrinsic time evolution of a quantum 
physical system, i.e. the time evolution of a state of a ``closed physical 
system'', isolated from all external influences, is described in the Hilbert 
space by (\ref{evol}) 
and is reversible.

The change described by (\ref{collax}) leads to the following problem: if 
$\rho=\rho(t)$ is time dependent and given by (\ref{evola}) 
(Schr\"{o}dinger picture) then $A$ and $P_{a_i}$ are time independent and 
$\rho^{\mathrm{after}}$ in (\ref{collax}) is at the same time as 
$\rho(t)$.  The collapse (\ref{collax}) 
is supposed to happen instantly and would not shed
any light on the question of a time arrow.  However,
realistically, every 
measurement takes time.  To avoid this inconsistency one 
considers the Heisenberg picture
~\cite{Hartl}:
 keep $\rho$ time independent $\rho=\rho(t_0=0)$ and equal to the
``initial'' state of (\ref{evol}), and take for the projection operators
the time evolved $P_{a_i}(t)=e^{iHt}P_{a_i}e^{-iHt}$.  Then in place
of (\ref{measb}) on has
\begin{equation}\label{twokind}
\rho^{\mathrm{eff}}(t_a)=P_a(t_a)\rho P_a(t_a),\ 
\mbox{with the condition that}\ t_0<t_a
\end{equation}
(or at least $t_a\geq t_0$, but $t_a=t_0$
would be the unrealistic case of instant measurement).  The change 
(\ref{twokind}) combines the two conventional changes of states in quantum
mechanics.  Its generalization for a sequence of observables, 
$A(t_a)$, $B(t_b)=\sum_j b_j P_{b_j}(t_b)$, etc. to a 
``history''~\cite{Hartl} 
$[\mathrm{\mathbf{P}}]=(P_{a}(t_a)P_{b}(t_b)\cdots P_{c}(t_c))$ is:
\begin{subequations}\label{timeo}
\begin{equation}\label{seque}
\rho^{\mathrm{eff}}(t_c) = \sum_{i,j,\ldots,k} P_{c_k}(t_c)\cdots P_{b_j}(t_b)
P_{a_i}(t_a)\rho(t_0)P_{a_i}(t_a)P_{b_j}(t_b)\cdots P_{c_k}(t_c).
\end{equation}
with the condition
\begin{equation}\label{cond}
t_0<t_a<t_b<\cdots<t_c
\end{equation}
\end{subequations}
Since the times ``after'' $t_a,t_b,\ldots,t_c$ cannot precede the 
``initial'' times $t_0<t_a,t_a<t_b\ldots$ one obtains the time ordering
(\ref{cond}).  In contrast to (\ref{measb}), where the change is to happen
instantly, for (\ref{twokind}) and (\ref{seque}) one can talk of a change in
time.  This time ordering follows from the idea that there is a knowable
initial (prepared) state $\rho$ from which the series of probabilities
\begin{subequations}\label{preprob}
\begin{equation}
{\cal P}(a, t_a) = \mathrm{Tr}\left(P_a(t_a)\rho \right)
=\mathrm{Tr}\left(P_a(t_a)\rho P_a(t_a)\right)
=\mathrm{Tr}\left(\rho^{\mathrm{eff}}(t_a)\right), 
\;\;\;\;\;\;t_a>t_0\label{preproba}
\end{equation}
\begin{multline}
{\cal P}(b, t_b; a, t_a)\\ = 
\mathrm{Tr}\left(P_b(t_b)P_a(t_a)\rho 
P_a(t_a)P_b(t_b)\right)=
\mathrm{Tr}
\left(P_b(t_b)\rho^{\mathrm{eff}}(t_a)\right), 
\;\;\;t_b>t_a>t_0,\label{preprobb}
\end{multline}
and in general
\begin{equation}
{\cal P}(c,t_c;\cdots ; b, t_b; a, t_a)  = 
\mathrm{Tr}\left(P_c(t_c)\cdots P_b(t_b)P_a(t_a)\rho 
P_a(t_a)P_b(t_b)\cdots P_c(t_c)\right)\label{preprobc}
\end{equation}
\end{subequations}
can be {\em pre}dicted (rather than retrodicted).  Though it may be an
oversimplification to think of $\rho^{\mathrm{eff}}(t)$ as the state of the
system after the measurement of the observable $P_a$, the 
$\mathrm{Tr}(\rho^{\mathrm{eff}}(t_a))$ are meaningful physical quantities,
namely the probabilities to observe with an apparatus the value $a$ for the
observable $A$ (cf.~Sect. \ref{Thegold} below).

The time ordering (\ref{cond})
expresses an arrow of time.  If one considers
an experiment performed on a quantum system in the laboratory with the
state $\rho$ prepared by the preparation apparatus (accelerator) and $P_a(t)$
the observable registered by the registration apparatus (detector), then 
the condition $t_0<t_a$ in (\ref{preproba}) expresses 
the preparation$\rightarrow$registration arrow of time 
that we mention above and
discuss in greater detail in Sect. \ref{Timesym} below, 
cf.~eq.~(\ref{detprob}).
This intuitively obvious time ordering and the arrow of time
it implies is the 
basis of the quantum mechanical irreversibility discussed in our
article.  The
initial time $t_0=0$ is the time at which the preparation of the state
$\rho$ is completed and
the registration of the observable $P_a(t)$ can begin.  The probability
 ${\cal P}(a, t_a)$ is the probability to register the observable 
$P_a(t_a)$ at the time $t_a$ in a prepared state $\rho$.  For example,
if $\rho$ describes the prepared state of a $K^0-\bar{K}^0$ system and
$P_a$ the projection operator on the $\pi^+\pi^-$ subspace then
${\cal P}(a, t_a)$ is the probability for decay of the neutral $K$ meson
into $\pi^+\pi^-$, which can be detected only at a time $t_a>t_0$.
Since the formation time scale of the neutral $K$ is ten orders of magnitude
less than that of the decay, the creation time $t_0$ is very precisely fixed.

The formula (\ref{preproba}) of the quantum mechanics of measured systems
has been generalized in 
\cite{Hartl} to (\ref{preproba}), 
(\ref{preprobb}),$\ldots$,(\ref{preprobc}) which have been understood 
to apply to the quantum 
theory of cosmology.  The probabilities are the probabilities of histories 
$P_a(t_a)$, $P_a(t_a)P_b(t_b)$,
$\cdots P_a(t_a)P_b(t_b) \cdots P_c(t_c)$ of such systems as our universe
as a whole.  The arrow of time expressed by the time ordering in
(\ref{seque}) and (\ref{cond})
would in this case be the time-asymmetry of the universe, i.e.
 the cosmological arrow of time.

Nothing has been said about the probabilities (\ref{preprob}) for
$t_a < t_0$, $\ldots$, $t_c<\cdots <  t_b < t_a < t_0=0$.  
In cosmology, where $t_0$ is the time of the big bang, we would not know.
In the decay process of a laboratory
experiment we could make the following conclusion: 
the detector does not click if it is switched
on before the decaying state has been prepared at $t=0$.  
It is therefore reasonable to 
assume that the probabilities (\ref{preprob}) are zero for these negative 
values of $t$.

As will be discussed in Sect. \ref{Conseq} below, in the Hilbert space theory
 (with no other further assumptions than 
$H$ is self-adjoint and semi-bounded) the 
probabilities of (\ref{preprob}) are identically zero (for all $t$) if they 
are zero on a negative time interval (actually on any set of non-zero measure).
Therefore one cannot implement the above program 
as a mathematical theory in the Hilbert space unless the probabilities
${\cal P}(a,t_a);{\cal P}(b,t_b;a,t_a)\cdots$ are
defined to be different from zero for all $t_a<t_0$, $t_b<t_a$, etc.  
The Hilbert space does not allow for a theory in which both energy 
{\em as well as} time have a lower bound (creation time).  
If one wants to have a 
quantum mechanical arrow of time, i.e. a mathematical formulation of 
 the obviously correct preparation$\rightarrow$registration arrow,
then one has to go beyond the Hilbert space formulation.  This
is the subject of the present paper indicated by the word rigged Hilbert space 
(also called Gelfand triplet) in the title.

Irreversibility in conventional quantum theory is always thought of as 
being due to external influences upon the non-isolated (``open'') quantum 
system.  The irreversible time evolution of open quantum systems 
is described by the master equation~\cite{Prigo}
\begin{equation}\label{masteq}
\frac{\partial \rho(t)}{\partial t} = L\rho (t)
\end{equation}
where $\rho(t)$ represents the state of the open system $S$, and the Liouville 
operator $L$ is given by
\begin{equation}\label{liouv}
L\rho(t)=-\frac{i}{\hbar}[H, \rho(t)] + \delta H (\rho)
\end{equation}
For $\delta H(\rho) = 0$, (\ref{masteq}) with (\ref{liouv}) is the
von Neumann equation whose solution is the 
reversible time evolution of the isolated quantum system given by 
(\ref{evola}).  Equation (\ref{masteq}) with (\ref{liouv}) is 
the standard equation for extrinsic irreversibility under the 
influences of an external reservoir $R$ (which could be, for example, a 
measuring apparatus) upon the 
system $S$.  The term $\delta H(\rho)$ represents some complicated external 
effects of the reservoir $R$ upon the quantum system $S$. 
 Under particular assumptions about the term $\delta H(\rho)$ the 
irreversible time evolution of $S$ can be shown to be described by a 
completely positive semigroup\footnote{According to A. Kossokowski 
(private communication) maps $\rho\rightarrow\rho(t)=\Lambda(t)\rho$ from the 
set density operators into itself could also describe ``some kind of 
intrinsic irreversibility'' if $\Lambda(t)$ is not completely positive, but
only positive.  But so far it is not clear whether these maps have any
physical meaning.} 
generated by a Liouvillian $L$~\cite{KKraus}):
\begin{equation}\label{semivol}
\rho(t)=\Lambda(t)\rho(0),\ \Lambda(t)=e^{Lt},\ \mbox{for}\ t\geq 0.
\end{equation}
This is the conventional semigroup evolution of open quantum 
systems 
(Sudarshan-\-Gorini-\-Kossakowski-\-Lindblad semigroup) \cite{ghir}.

This time evolution described by a Liouvillian $L$, where $L\rho$ is not just 
the commutator with the Hamiltonian of the system, $\frac{1}{i\hbar}[H,\rho]$,
has also been applied to the decaying $K^0$-system.  It can evolve a
pure state $\ktbr{\phi}{\phi}$ into a mixture and has been called
non-quantum mechanical~\cite{Huet}.

The `Irreversibility' that we mean in the title is
not the non-quantum mechanical irreversibility described by (\ref{semivol})
with (\ref{liouv}) and it is not the increase of the von Neumann entropy 
(\ref{entrop}) due to the collapse axiom (\ref{collax}) and 
(\ref{mixt})\footnote{The measurement process which changes $\rho$ to $\rho'$
is a scattering process of a microsystem on a macrosystem (``measurement 
scattering''~\cite{Ludwig}) and does not actually fulfill the idealized 
measurement axiom (\ref{collax}).  Every scattering experiment possesses
 an arrow of time; preparation must always precede registration.  The arrow of
time for the change of state due to measurement is thus a consequence of the 
preparation$\rightarrow$registration arrow, G. Ludwig, private 
communication.``I know of no `other' time arrow than the 
preparation$\rightarrow$registration arrow.''}.
And again, the `Semigroup' that we mean
is not the semigroup (\ref{semivol}) 
generated by the Liouvillian (\ref{liouv}).
Our semigroup is the semigroup generated by the Hamiltonian of the 
isolated quantum system:
\begin{equation}\label{isovol}
e^{-iH^{\times}t},\ t\geq 0 (=t_0),
\end{equation}
where $H^{\times}$ is the extension of a self-adjoint (semi-bounded) Hilbert
space operator $\bar{H}$.  The semigroup (\ref{isovol}) is obtained from
the same time-symmetric dynamical equations (the Schr\"{o}dinger equation
(\ref{evolc}) or the von Neumann equation) from which one obtained 
the unitary group
in (\ref{evola}) and (\ref{evolb}) by integration.  
However, whereas the unitary
group evolution, (\ref{evola}) and (\ref{evolb}), is obtained if one
requires that the set of $\phi$ in (\ref{evolc}) are elements of the Hilbert
space ${\cal H}$, the semigroup evolution (\ref{isovol}) is obtained
from (\ref{evolc}) if 
one requires that the set of allowed solution $\phi$ in (\ref{evolc}) are 
elements of
a space $\Phi^{\times}_+$ that extends
the Hilbert space ${\cal H}$ to a Gelfand triplet 
$\Phi^{\times}_+\supset{\cal H}\supset\Phi_+$.  
The arrow of time expressed by the
time evolution semigroup (\ref{isovol}) 
thus arises mathematically from time-symmetric 
dynamical equations solved with time-asymmetric boundary conditions.
  These time-asymmetric boundary conditions 
which follow from the preparation$\rightarrow$registration arrow of time,
cannot be mathematically formulated in the Hilbert space; in ${\cal H}$ the 
equations (\ref{evolc}) 
always integrate to the one-parameter group (\ref{evolb}).

The semigroup time evolution (\ref{isovol}) describes irreversible time
evolution on the microphysical level (if one interprets the solutions of the 
Schr\"{o}dinger equation in the extension $\Phi^{\times}$ of $\bohmHS$
as microphysical states).  Examples of isolated 
(closed, autonomous) microphysical systems
with irreversible time evolution abound in the real world.
Reversible time
evolution is a feature of only a minority of closed microphysical systems,
e.g.~the ground states of molecules and nuclei and the few stable elementary
particles.  There is a much larger number of decaying states and resonances
(excited states of molecules and nuclei, weakly decaying elementary particles,
hadron resonances) which are not less isolated than ground states;
their time evolution is irreversible (which is sometimes mentioned in
textbooks~\cite{C-T}).  

It has been argued that a decaying state or resonance is something 
complicated.  The difficulty is that in Hilbert space $\bohmHS$ one cannot
find a simple initial condition for it.
However, in its extension $\Phix_+$ there is a simple (pure) 
initial state $\rho_i=\kb{E_R-i\Gamma/2}{E_R-i\Gamma/2}$ at the time $t_0$,
the creation time of the resonance (e.g.~the time when the $K^0$ is leaving
the proton target in $\pi^- p\rightarrow \Lambda K^0$ of the many 
experiments on neutral Kaon decay and CP-violation).  The decaying state
vector $\psi^G=\kt{E_R-i\Gamma/2}\in\Phix_+$ and is a 
generalized eigenvector of the
self-adjoint Hamiltonian $H$ with complex eigenvalues 
$z_R=E_R -i\Gamma/2$ associated to the second sheet pole of the 
$S$-matrix~\cite{PhysicaA}.
  (A resonance is
actually described by a pair of poles $z_R=E_R\mp i \Gamma /2$, a pair of
vectors $\psi_\mp^G$ in a pair of spaces $\Phi^\times_\pm$ evolving by a
pair of semigroups, cf.~Sect.\ \ref{Semigr}).

The semigroup time evolution of the Gamow vector,
 is derived as~\cite{PhysicaA}:
\begin{equation}\label{psig}
e^{-iH^{\times}t}\psi^G=e^{-iE_Rt}e^{-\frac{\Gamma}{2}t}\psi^G,\ 
\mbox{for}\ t\geq 0\ \mbox{only}.
\end{equation}
This is 
a mathematical consequence of the (mathematical formulation of the) 
preparation\-$\rightarrow$\-registration arrow of time and vice versa (derived
using the Paley-Wiener theorem).  The time-asymmetry 
of resonances expressed by the semigroup
(\ref{psig}) is not the irreversibility of quantum mechanical measurements;
resonances evolve in time and decay without being ``looked at''.  
But it is identical with the arrow of quantum cosmology if (\ref{twokind})
 and (\ref{seque}) are applied to 
the initial state of the universe.  The irreversibility 
that we mean is the asymmetric time evolution on the quantum level whose
basis is the causality principle, which in turn can be inferred from a
very special state of the universe at the beginning of time.

\section{Introduction---Dirac Kets}

\footnotesize
\begin{quote}
``Physical causality can be traced directly to the existence of a simple
initial condition of the universe.  But how does that initial condition enter
into the theory?''

---Murray Gell-Mann\footnote{p.~216 of M.Gell-Mann, The Quark and the Jaguar,
(W.H. Freeman, New York, 1994).}
\end{quote}
\normalsize

A mathematical theory of physics cannot be decuded from experiments alone;
it will necessarily have to involve some idealizations.  Mathematicians like
to work with mathematical structures that are complete---algebraically
complete and, in particular, topologically complete.  Only with such
structures can they prove existence theorems.  Experiments can never be
complete in this sense; physicists have only a finite amount (albeit 
very large sometimes) of experimental data with which to work.
Mathematicians generalize to infinity through various means.
The topology\footnote{\label{dictfn} Webster's dictionary
 gives three definitions of the
word topology: (1) the study of those properties of geometric forms
that remain invariant under certain transformations, as bending,
stretching, etc.; (2a) the study of limits in sets considered as
collections of points; (2b) a collection of open sets making a given
topological space.  Physicists usually associate definition 1. with
the word topology, we here use only definition (2).  Mathematical
structures are combinations of three basic kinds of structures:
algebraic, topological and ordering~\cite{topodef}.  The rigged
Hilbert Space is the completion of the same linear (algebraic) space 
equipped with three
different topologies.} of a linear space defines the 
meaning of convergence of
infinite sequences and the topological completion of a space is the 
appendage to the linear space 
of the limit elements of all infinite convergent (to be 
precise, Cauchy) sequences.
A finite number of experiments, even if the number is arbitrarily 
large, cannot tell how the convergence of infinite sequences of states
should be defined. Therefore the choice of topology cannot be ``deduced'' 
from experiments and only the overall success of a mathematical 
theory can show that
one topology is preferable to another.

This paper explores different choices of topological 
completions for the spaces of states and observables.  One is given by
von Neumann's Hilbert space completion.  The others---actually a multitude of
topologies, one for each particular kind of quantum physical system---are 
conjectured from Dirac's formalism.

Dirac~\cite{Dirac} introduced such elements as bras \br{E} and \br{x},
kets \kt{E} and \kt{x} and an algebra of observables generated by such
fundamental operators as the Hamiltonian $H$, momentum $P$ and
position $Q$.  Of these kets he demanded that they were eigenvectors
\begin{eqnarray}
\label{evect}
H\kt{E} & = & E\kt{E}\nonumber\\ Q\kt{x} & = & x\kt{x}
\end{eqnarray}
and that they formed a complete basis system, i.e. that every vector
$\phi$ could be written as
\begin{subequations}\label{expans}
\begin{equation}\label{expansa}
\phi = \sum_{n=1}^{\infty}\dkt{E_{n}}\dbk{E_{n}}{\phi} + \int
dE\,\kt{E}\bk{E}{\phi},
\end{equation}
\begin{equation}
\phi = \int dx\, \kt{x}\bk{x}{\phi}
\end{equation}
or
\begin{equation}
\bk{x'}{\phi} = \int dx\,\bk{x'}{x}\bk{x}{\phi},
\end{equation}
\end{subequations}
where \dbk{E_n}{\phi}, $\bk{E}{\phi}\equiv\phi(E)$ and 
$\bk{x}{\phi}\equiv\phi(x)$ 
(the wave function) were thought of as scalar products, 
$\phi(x)=\ip{\kt{x}}{\phi}$.

These intuitive constructions were the results of Dirac's unconfined
vision, but not well-defined mathematical objects.  A mathematical
structure to envelope all Dirac's calculative tools was not available
at that time.  The delta function $\bk{x'}{x}=\delta(x'-x)$ inspired
the development of distribution theory by Schwartz~\cite{Scwartz} and
the \kt{x}, \kt{E} inspired the creation of the rigged Hilbert space
by Gelfand and his school~\cite{Gelfand}[18a].

In the following discussion, we leave aside the proper eigenvectors 
$H\dkt{E_n}=E_n\dkt{E_n}$ with discrete eigenvalues $E_n$,  which 
are conventionally 
negative $E_n=-|E_n|$, and the energy continuum starts at
zero so that the integral in (\ref{expansa}) extends from $0$ to $\infty$ 
(there can also be discrete eigenvalues in the continuous spectrum and then 
these $\dkt{E_n}$ with $E_n>0$ would also be included in the discrete sum of 
(\ref{expansa})).  In the $S$-matrix they correspond to poles on the
negative real axis of the first sheet (bound state poles).  Both the discrete 
and continuous values for the energy must be bounded from below (stability
of matter) and by convention the lower bound for the energy continuum is 
chosen to be zero.

The discrete basis vectors describe stationary states: $W=\sum_n w_n
\dkb{E_n}{E_n}$.  The linear space $\Psi^{\mathrm{disc}}$ spanned by 
these eigenvectors of the discrete spectrum could also be completed into 
a Hilbert space $\bohmHS^{\mathrm{disc}}$ (or a rigged Hilbert space) which is
orthogonal to the Hilbert space $\bohmHS$ of the continuous spectrum.  In the 
following, we will only consider the linear space and their completions
$\Psi$, $\Phi$, $\bohmHS$ which belong to the absolutely continuous spectrum.  
The quantum mechanics of stationary states is not affected by our 
considerations here.  Thus $\bohmHS$ in this paper denotes the Hilbert space of
the absolutely continuous spectrum, often denoted as $\bohmHS^{\mathrm{a.c.}}$
in the literature.

\section{\label{Hilbert} Hilbert Space (HS) and Rigged Hilbert 
Space (RHS) Formulation of Quantum Mechanics} 

\footnotesize
\begin{quote}
``I would like to make a confession which may seem immoral: I do not believe
in Hilbert space any more.''

---John von Neumann\footnote{In a letter to G.~Birkhoff,
quoted in G.~Birkhoff, Proceedings of Symposia in Pure Mathematics, Vol.~2, 
ed.~R.P.~Dilworth, (American Mathematical Society, Providence, Rhode Island, 
1961), p.~158.  The letter is date Nov.~13, and Birkhoff believes the year was
1935.}
\end{quote}
\normalsize

The first attempt at a rigorous mathematical theory for quantum
physics was provided by Weyl~\cite{Weyl} and von Neumann~\cite{vonNeu}
using the mathematics that was available at that time, the Hilbert
space (HS).  The HS is the completion of a linear scalar product space
(also called pre-Hilbert space) with respect to the
topology\footnote{cf.~footnote (\ref{dictfn}).} given by the norm
$||\phi||=\sqrt{\ip{\phi}{\phi}}$.

This norm topology is one of many possible choices and
cannot be deduced from physical observations of quantum mechanical
states and the observables represented by the operators.
The HS properties are:
\begin{enumerate}
\item The wave functions $\psi(x)\equiv\bk{x}{\psi}$ representing the
probability $\norm{\bk{x}{\psi}} \Delta x$ to detect the particle
state $\psi$ within the position interval $\Delta x$ or the energy
wave function $\phi(E)\equiv\bk{E}{\phi}$ representing the the energy
distribution in a particle beam are given in the HS by a {\em class}
of Lebesgue square integrable functions \{$\psi(x)$\} or $\{\phi(E)\}$
that differ on a set of Lebesgue measure zero.
\begin{eqnarray}
\psi\in\bohmHS & \Leftrightarrow & \{\psi(x)\}\in {\cal L}^2
({\Bbb R})\nonumber\\ \phi\in\bohmHS & \Leftrightarrow &
\{\phi(E)\}\in {\cal L}^{2}({\Bbb R})\label{class}
\end{eqnarray}
In a scattering experiment, $\norm{\psi(x)}$ represents the detector
size, location and efficiency and the $\norm{\phi(E)}$ represents the
energy resolution of the accelerator.  Unlike the classes of
(\ref{class}), the resolution of an experimental apparatus is always
given by a single {\em smooth} function, e.g.~$\psi(x)\in{\cal S}$
(Schwartz space), and not by a class
$\{\psi_{1}(x),\psi_{2}(x),\psi_{3}(x),\ldots\}$ of functions which
can vary wildly on any set of Lebesgue measure zero, e.g.~on the set
of all rational numbers.

One can always choose the one smooth function
$\psi(x)\in{\cal S}$ in this class $\{\psi(x)\}\in{\cal L}^2({\Bbb R})$,
 but the set of smooth functions (or more precisely
the set of classes of Lebesgue square integrable functions containing a 
smooth function $\psi(x) \in {\cal S}$) is not
HS-complete.  For the HS-completeness, one needs the Lebesgue
integral~\cite{Reed}.

The notion of Lebesgue integrability is physically counterintuitive.
Physicists make a finite number of measurements and interpolate smoothly 
between the experimental data.  They never compute Lebesgue integrals, but 
calculate Riemann integrals using the smooth functions associated to their 
experimental apparatus.

\item The most fundamental commutation relations, such as $[P, Q]=-i1$
of momentum $P$ and position $Q$, imply that these observables cannot
be represented by continuous (and hence bounded) operators in HS.
Therefore $P$ and $Q$ have only a limited domain of definition in
${\cal H}$.  Physicists work with operators that one can add and
multiply (Dirac's algebra of observables) and not with operators that
cannot be defined on the whole space.  Often such an algebra is the
enveloping algebra of a symmetry group which represents physical
transformations, for example, of the registration apparatus (detector)
relative to the preparation apparatus (accelerator).
One can always find a subspace $\Phi\subset\bohmHS$ on which
these operators are defined (e.g.~the space of differentiable vectors
or the space of analytic vectors).  Then one can define a
topology\footnote{cf.~footnote (\ref{dictfn}).} with respect to which these
operators are continuous.  The space $\Phi$ cannot be a Hilbert 
space unless it is a
finite-dimensional representation space of a compact group. However,
such a subspace $\Phi$ can be represented a space of the smooth functions,
in many cases (e.g.~for $Q$ or $P$) by the space
${\cal S}$:
\begin{equation}
\phi\in\Phi\Leftrightarrow\phi(x)\in{\cal S}.
\end{equation}

\item The HS does not contain eigenkets like \kt{E} and \kt{x} or the
bras \br{E} and \br{x}, with the properties (\ref{evect}) and
(\ref{expans}).  Physicists use them however, e.g.~as scattering
states, and (\ref{expans}) is the fundamental relation for
computations in quantum theory.  The kets provide an opportunity to
describe states of single microphysical systems.

In von Neumann's formulation with $\phi,\psi\in\bohmHS$, $\phi$ represents
the state of an ensemble and \kb{\psi}{\psi} represents the observable
on an ensemble.  The probability to measure the observable
\kb{\psi}{\psi} in the ensemble $\phi$ is \norm{\ip{\psi}{\phi}}.  The
vectors $\psi$ and $\phi$ are defined by the experimental apparatuses.
For example, if $\phi$ is the idealized representation of a beam of
microphysical particles prepared by an accelerator (idealized, because
real accelerators prepare mixtures $\sum\kb{\phi_{i}}{\phi_{i}}=W$),
then
\begin{equation}
\frac{\int_{\Delta
E}dE\norm{\phi(E)}}{\int_{\mathrm{all}\,E}dE\norm{\phi(E)}}
\end{equation}
is the fraction of this large number of particles that have energy in
the interval $\Delta E$.  One never talks of a single microphysical
particle but always of a large number, an ensemble.  (Equivalently one
can talk of an ensemble of experiments.)  The notion of a single
microsystem or a state of a single microsystem does not exist in the
standard HS formulation; $\phi$ describes an ensemble and
$\phi(E)\in\{\phi(E)\}\in {\cal L}^{2}({\Bbb R}_{+})$ 
represents the energy
distribution of that ensemble for which the physicists always choose
a smooth function $\phi^{smooth}(E)$.  Physicists do not work with 
the {\em class} of 
Lebesgue square integrable
functions $\{\phi_1(E), \phi_2(E),\ldots,\phi^{smooth}(E)\}$ 
which contains the smooth function.  They cannot even 
isolate experimentally those classes 
of functions which do not contain a smooth 
function $\phi(E)=\phi^{\mathrm{smooth}}(E)\in{\cal S}$, i.e. the
elements of ${\cal L}^2$ which are not elements of ${\cal S}$.  
That means the class $\{\phi(E)\}\in{\cal L}^{2}({\Bbb R}_{+})$ 
which does not contain a $\phi^{smooth}(E)$ has
no physical meaning.

The probability amplitude \bk{\psi}{\phi} of the physicist is therefore 
given by
\begin{subequations}\label{prob}
\begin{eqnarray}\label{rieint}
\bk{\psi}{\phi} & = &
\sum_{i=1}^{\infty}\bk{\psi}{i}\bk{i}{\phi}\nonumber =
\int_{\mathrm{Riemann}} dE \bk{\psi}{E}\bk{E}{\phi} = \int_{\mathrm{Riemann}}
 dx \bk{\psi}{x}\bk{x}{\phi}\\ & = & \int\int_{\mathrm{Riemann}} dx\,dE
\bk{\psi}{x}\bk{x}{E}\bk{E}{\phi}
\end{eqnarray}
where $\kt{i}$, $i=1,2,\ldots,n,\ldots$, is some discrete basis system,
where the wave functions $\psi(x)\equiv\bk{x}{\psi}$ and 
$\phi(E)\equiv\bk{E}{\phi}$ are smooth functions and where the integrals
are Riemann.  The probability amplitude in the HS formulation is calculated
as:
\begin{equation}\label{lebint}
\bk{\psi}{\phi}=\int_{\mathrm{Lebesgue}}dx\,\psi^*(x)\phi(x) =
\int_{\mathrm{Lebesgue}}dE\,\psi^*(E)\phi(E).
\end{equation}
\end{subequations}
The integrals are Lebesgue and the set of mathematical wave functions,
i.e. the classes $\{\psi_1(x), \psi_2(x),\ldots\}\in{\cal L}^2$ in
(\ref{lebint}), also includes elements that contain no $\psi^{smooth}(x)\in
{\cal S}$.  Since the physicist only deal with smooth energy distributions
\norm{\phi(E)} or smooth position distributions \norm{\psi(x)} due to
the capabilities attributed to their apparatuses, it is 
of no advantage to choose some arbitrary Lebesgue integrable
$\phi(E)$ or $\psi(x)$ as the tool for the calculation of
(\ref{lebint}).  A much ``more practical method of computation'' and the
method that a physicist always uses is to choose for the
$\phi(E)\in\{\phi(E)\}$ the
$\vpe=\phi^{\mathrm{smooth}}(E)\in{\cal S}$.  Then the Lebesgue
integrals in (\ref{lebint}) become Riemann integrals and are identical with 
(\ref{rieint})\footnote{cf.~footnote (\ref{prereq}) of Sect. (\ref{sechalf})
for the prerequisites on the operators $H$, $Q$ to make this possible.}.  
The other integrals in (\ref{lebint}), for which 
$\{\psi_1(x),\psi_2(x),\ldots\}$ does not contain a $\psi^{smooth}(x)$,
have no observational counterpart anyway 
and are therefore physically useless.  They
have to be included in (\ref{lebint}) because only with them is the HS a 
topologically complete space.

Only for $\bk{E}{\phi}=\phi^{smooth}(E)\in{\cal S}$, and only if
one restricts oneself to the integrals of (\ref{rieint}), can one give to the
symbol \br{E} in \bk{E}{\phi} a separate mathematical meaning by
interpreting \bk{E}{\phi} as the value of an anti-linear functional
\br{E} at the element $\phi\in\Phi$.  This means one can define \br{E}
as an element of $\Phi^{\times}$, the space of continuous anti-linear
functionals over $\Phi$.  Since this is only possible for the smooth
\vpe, i.e. for the $\phi\in\Phi$, not for all elements of the HS, we
obtain for the set \Phix\ a larger set than the
set of anti-linear continuous functionals over \bohmHS, since from $\Phi\subset\bohmHS$
 it follows $\bohmHS^{\times}\subset\Phix$.  Using the Frechet-Riesz
theorem~\cite{Reed} we then identify $\bohmHS=\bohmHS^{\times}$ and obtain a
triplet of spaces, the Gelfand triplet or rigged Hilbert space (RHS):
\begin{equation}
\Phi\subset\bohmHS=\bohmHS^{\times}\subset\Phi^{\times},
\end{equation}
where $\kt{E}\in\Phi^{\times}$.

The desire for a ``more practical method of computation'' using
Riemann integrals and observable quantities only
 has thus led to a choice of \bk{E}{\phi} which
allowed us to give a mathematical meaning to the kets \kt{E} and
\kt{x}.  These kets lie outside the HS, whose elements (and therewith
also the $\phi\in\Phi\subset\bohmHS$) describe ensembles of microphysical
systems.  These new vectors, the kets \kt{E}, are available for a
physical interpretation that goes beyond the ensemble interpretation.

Though one can only observe the probabilities (\ref{rieint}) measured
with macroscopic apparatuses, it is intuitively attractive to imagine
that the effect which the preparation apparatus causes on the
registration apparatus is carried from one to the other by some
physical entities.  These objects, by which the preparation apparatus
acts on the registration apparatus (i.e. the carriers of the action),
are imagined to be the microphysical systems.  Though one cannot see
them directly, every physicist believes in them, e.g.~believes that
the track in a cloud chamber is caused by a single particle.  In HS
there is nothing that can describe them, but in RHS the kets of
$\Phi^{\times}$ may.

Whereas the physical entities connected with an experimental
apparatus, like the states $\phi$ defined by the preparation apparatus
or the property $\psi$ defined by the registration apparatus, are
assumed to be elements of \bohmHS, or as described above even of $\Phi$,
the imagined entities connected with microphysical systems do not have
to be in \bohmHS\ or $\Phi$ because the energy distribution for a
microphysical system does {\em not} have to be a measurable or even a
continuous, infinitely differentiable, rapidly decreasing function of
the physical values of $E$, like the \bk{E}{\psi} describing the
energy resolution of the detector, or the \bk{E}{\phi} describing the
energy distribution of the beam.  For the hypothetical entities
connected with single microphysical systems one can use Dirac's kets
\kt{E_0} with the energy distribution
$\bk{E}{E_0}\propto\delta(E-E_0)$.  The energy eigenkets \kt{E,
\Omega} (or momentum kets \kt{\mathbf{p}}, $E=\mathbf{p}^{2}/2m,
\Omega=\mathbf{p}/|\mathbf{p}|$) represent then the microphysical
states of momentum \mom.  These are not states one can prepare with a
macroscopic apparatus, but something that the physicist imagines as
single microsystems, something that can be associated with a cloud (or
bubble) chamber track.
\end{enumerate}

Summarizing, the HS is too big if one only admits quantities
associated with macroscopic apparatuses because apparatuses have
smooth energy distributions, $\vpe\in{\cal S}$ and not every
element of $L^{2}({\Bbb R})$ describes an energy distribution of an
apparatus, i.e., a physically preparable state $\phi\in\Phi$.
However, if one also wants to describe single microphysical systems,
then the HS is too small, because microphysical states like Dirac's
scattering states \kt{\mom} cannot be represented in \bohmHS.  Neither can
Gamow's decaying states \kt{E - i\Gamma/2} representing microphysical
systems with well defined resonance energy $E$ and lifetime
$\hbar/\Gamma$ be represented in \bohmHS.  The Gamow kets are the
principal objects of this paper.

Usually von Neumann's formulation of quantum mechanics entails some
further idealizations in addition to the choice of the HS which one
may or may not want to make, like:
\begin{list}
{\alph{vnstuff}}{\usecounter{vnstuff}}
\item The one-to-one correspondence between the set of states
(equivalence classes of preparation apparatuses) and the set of
statistical operators $W$ in \bohmHS, or the one-to-one correspondence
between pure states and elements of \bohmHS.  This is already an
exaggeration on the pre-Hilbert space level, since not every finite
superposition of physical state vectors which will represent a
physical state as has recently been emphasized in the discussion of
decoherence~\cite{Zurek}.
\item The one-to-one correspondence between the original set of
observables and the set of self-adjoint operators $A$ in \bohmHS.
\item The axioms of the idealized measurements (collapse of the wave
function). Realistic experiments usually do not even attempt to
fulfill the condition of an ideal measurement (of first or second
kind) and according to present views this postulate is not needed,
since the Schr\"{o}dinger equation and a measurement scattering
process fully describe the measuring act~\cite{Ludwig,Kraus}.
\end{list}
Since these additional idealizations are not of direct relevance for
our discussion of quantum mechanical irreversibility and have no
bearing on the choice of the most suitable topology for the scalar
product space, we shall not discuss them any further here.

After having discussed the basic features of the HS formulation we
shall now discuss the RHS formulation of quantum 
mechanics~\cite{Bohm,Roberts,B-n-G}, which has
only become possible after the new mathematics of distributions and
linear topological spaces had been introduced 30 years after von
Neumann's HS.

The RHS formulation is also a mathematical idealization of the
structure that one can deduce from physical observations.  This
idealization provides the fundamental properties (\ref{evect}) and
(\ref{expans}) required by Dirac's formalism.  It restricts the
allowable vectors for the description of entities defined by the
experimental apparatus to the subspace $\Phi\subset\bohmHS$; $\Phi$ is
complete with respect to a different topology (meaning of convergence)
than \bohmHS.  But this is of no importance to physics because physics
will not use all elements of $\Phi$; important is that this topology
is such that (\ref{evect}) can be mathematically defined (see Sect.\ 
\ref{Froma} below) and (\ref{expans}) can be proved.
\begin{table}
\centering
\begin{tabular}{|p{0.29\textwidth}|p{0.29\textwidth}|p{0.29\textwidth}|}
\hline HS formulation of theoretical quantities & Experimental
quantities & RHS formulation of theoretical quantities\\ \hline\hline
\multicolumn{3}{|c|}{Preparation Apparatus}\\
\multicolumn{3}{|c|}{defines the}\\ \hline Density operator $W$ in
${\cal H}$ & Physical states (mixtures)& Density operator $W$ in
$\Phi_-$\vspace{2pt}\\ \hline $\phi$ element of ${\cal H}$ or
projection operator \kb{\phi}{\phi} & pure states & $\phi$ element of
$\Phi_-\subset{\cal H}$\\ \hline\hline
\multicolumn{3}{|c|}{Registration Apparatus}\\
\multicolumn{3}{|c|}{defines the}\\ \hline Unbounded linear operator
$A$ in ${\cal H}$ & Physical observables & Element of an algebra of
$\tau_{\Phi}$-continuous operators\vspace{2pt}\\ \hline Projection
operator $\Lambda$ or self-adjoint operators $F$ with $0\leq F \leq 1$
in \bohmHS & & Projection operators or positive operators in $\Phi_+$\\
\hline $|\psi\rangle\langle\psi|$ one dim. projector in ${\cal H}$
&Yes-no observables (property or proposition) &
$|\psi\rangle\langle\psi|$ with $\psi\in\Phi_+\subset\Phi$\\ \hline
\end{tabular}

\vspace{4pt}The spaces $\Phi_+$ and $\Phi_-$, with $\Phi=\Phi_- +
\Phi_+$; $\Phi_-\cap\Phi_+\neq 0$, will be defined in Sect.\ \ref{Froma}.
\caption{Comparison of HS and RHS description of physical
quantities.}
\end{table}

The space $\Phi$ is specific to the particular quantum physical system
considered, and the topology of the space $\Phi$ is defined such that:
\begin{enumerate}
\item The algebra of observables of the quantum physical system is an
algebra of continuous operators in $\Phi$.

\item The Dirac basis vector expansion (\ref{expans}) holds as a
theorem, the Nuclear Spectral Theorem.
\end{enumerate}
The distinction between the RHS and the HS formulations of quantum
mechanics is summarized in Table I.

For the quantum mechanics of stationary states and reversible
processes (using no more than Dirac kets), the HS formulation and the
RHS formulation lead to experimental predictions which are only
imperceptibly different.  The calculations can be written in terms of
elements which can be defined with either the HS or the RHS.  The
standard HS quantum mechanics is just an approximate sub-theory of RHS
quantum mechanics.  This similarity reflects the similarity of von
Neumann's HS formulation and Dirac's incomplete bra and ket formalism.
The RHS formulation of quantum mechanics makes the Dirac formalism
rigorous and provides a more ``practical method of computation'' in
the domain where both theories overlap, e.g.~by allowing the use of
Riemann integrals in computations like (\ref{prob}) rather than
Lebesgue integrals.

But with the new mathematical language that the RHS provides one can
speak and think new physics.  The microphysical scattering states
(hyperbolic orbits of the corresponding classical system) described by
the Dirac kets \kt{E} gave already an indication of this new domain.
The really new physics of the RHS formulation is the microphysical
irreversibility which is exemplified by the semigroup time evolution
of the Gamow kets which represent decaying states or resonance states.
This irreversible semigroup time evolution was unthinkable in the old
HS formulation, where decaying states had complicated, problematic
features, as we shall discuss next.

\section{\label{Conseq} Consequences of the HS Formulation and
Some Views on Time-Asymmetry in Quantum Mechanics}

Most computations in quantum theory do not use the completeness property of
the HS; they work only with properties of a pre-Hilbert space.  But
there are some general results that one obtains from HS-completeness which 
unveil the problems of the HS formulation of quantum decay:

\begin{enumerate}
\item There is no vector that obeys the exact exponential decay
law~\cite{Khalfin}.  Mathematically stated, there is no
$\phi\in{\cal H}$ whose survival probability
\begin{displaymath}
{\cal P}_{\mathrm{s}}\equiv\norm{\ip{\phi}{e^{-iHt}\phi}}
\end{displaymath}
has the property
\begin{equation}
{\cal P}_{\mathrm{s}}=e^{-\Gamma t}.\label{surv}
\end{equation}

Although this ``deviation from the exponential law'' is unobserved,
the theoretical prediction of it has led some to infer that exact
exponential decay does not exist in nature, instead of suspecting a
flaw in the mathematical idealization of the HS.  Since magnitudes
cannot be predicted from only mathematical properties, one can argue
that this deviation from the exponential law is smaller than any
experimental error.  Thus these deviations could always be smaller
than what can be experimentally ruled out and therefore cannot be 
tested%
\footnote{
Probabilities like ${\cal P}_{\mathrm{s}}$ are
always observed as ratios of (preferably large) integers
$N(t)/N(0)$ and not as real numbers like $e^{-\Gamma t}$, therefore any
discussion of deviations from the exponential law is futile as long as the 
deviations are not predicted with a magnitude that can be compared with 
$1/\Gamma$. }.
In spite of the 
untestibility of these mathematical deviations, alternate explanations 
for the observed exponentiality have been proposed.  For example,
the observed exponential behavior has been ascribed to the influence of the
environment, to the measurement process or to both~\cite{Fonda}.

Physicists usually demand even more of a decaying state than
(\ref{surv}).  As envisioned by Gamow~\cite{Gamova}, physicists would
like to describe a decaying state by an eigenvector
$\psi^{G}\equiv\kt{E_{R}-i\Gamma/2}$ of the self-adjoint Hamiltonian
$H$ with complex eigenvalue ($E_{R}-i\Gamma/2$), i.e.
\begin{equation}
H\gktp = (E_{R} - i\Gamma/2)\gktp,\label{gktpevect}
\end{equation}
and the exponential time evolution
\begin{equation}
\psi^G(t) =
e^{-iHt}\psi^{G}=e^{-iE_{R}t}e^{-\frac{\Gamma}{2}t}\psi^{G}\label{gktpevol}
\end{equation}
but such vectors do not exist in the HS.

Empirically, stable particles are not considered qualitatively
different from quasi-stable particles, but are only quantitatively
different by a zero or very small value of $\Gamma$.  A particle
decays if it can decay and it is stable if selection rules for some
quantum numbers prevent it from decaying.  Both stable and decaying
states have been described in elementary textbooks~\cite{Frauen}, in
successful phenomenological or effective theories~\cite{LOY} and in
tables of experimental data~\cite{PartData}, as autonomous entities
characterized by $E_{R}$ and $\Gamma$, where $\Gamma$ can sometimes be
equal to zero.  A vector like $\psi^G$ fulfilling (\ref{gktpevect})
and (\ref{gktpevol}) would have the suitable properties.

Though there are no $\psi^G$ in the HS, the RHS contains these Gamow
vectors $\psi^{G}$.  They are in a conjugate space $\Phi_{+}^{\times}$
of a RHS, $\Phi_{+}\subset\bohmHS\subset\Phi_{+}^{\times}$.

\item Decay probabilities in the HS theory are identically zero.
The probability for the transition from a state
$\psi(t)=U^{\dag}(t)\psi=\exp{(-iHt)}\psi$ into the decay products
described by the subspace $\Lambda\bohmHS\subset\bohmHS$, where $\Lambda$ is
the projection operator on the subspace of decay products (or
$\Lambda$ is a positive operator), is in quantum theory given by
\begin{equation}
{\cal P}(t)=\mathrm{Tr}(\Lambda\kb{\psi(t)}{\psi(t)})
=\mate{\psi}{e^{iHt}\Lambda
e^{-iHt}}{\psi}.\label{proj}
\end{equation}
This is the probability to detect the observable $\Lambda$ in the
state \kb{\psi(t)}{\psi(t)}.  The Hamiltonian $H$ is always assumed to
be self-adjoint and semi-bounded $H\geq 0$, (the condition for the
stability of matter).  The decay of a prepared quasi-stationary state
is assumed to start at finite time, $t>t_2>-\infty$, which is
mathematically formulated as:
\begin{equation}
\int^{t_{2}}_{t_{1}}\mate{\psi(t)}{\Lambda}{\psi(t)}dt = 0\label{int}
\end{equation}
for some $t_{1}$ such that $-\infty\leq t_1 <t_2$
($t_{2}$ is usually chosen to be $t_2=0$).
With these assumptions one can show~\cite{Heger} 
\begin{equation}\label{impossdec}
{\cal P}(t)\equiv 0\ \mbox{for all}\ t\ (\mbox{precisely,
almost all}\ t)\ \mbox{and for {\em every}}\ \psi\in\bohmHS.
\end{equation}
This means that in HS the transition
probability ${\cal P}(t)=\mathrm{Tr}(\Lambda(t)\rho)$ of 
any state $\rho=\sum_iw_i\ktbr{\psi_i}{\psi_i}$ 
is identically zero if the transition starts at a finite
time.  In particular there are no decaying states in \bohmHS.  
This mathematical result is not usually interpreted as a
deficiency of the HS idealization but as being due to some problems
with causality~\cite{Buch}.

The way out of this dilemma is shown in Sect.\ \ref{Thegold}.
One replaces $\psi$ of the HS in (\ref{proj}) by the Gamow vector
$\psi^{G}$ of the RHS, which is not in the HS.  Then the
transition probability ${\cal P}(t)$ can be shown to be nonzero and
exponentially approaching unity, i.e., ${\cal P}(t)\propto
[1-\exp{(-\Gamma t)}]$, for $t>0$.  

\item In the HS formulation of quantum mechanics the symmetry
transformations (e.g.~Galilean transformations, Poincar\'{e}
transformations) are described by a unitary, projective group
representation in \bohmHS.  This implies that the time evolution is
unitary and reversible and given by $U^{\dag}(t)=\exp{(-iHt)},
-\infty<t<\infty$.

Instead of recognizing that this may be a property of the mathematical
idealization imposed by the HS, the widespread conclusion was that
quantum mechanical irreversibility of isolated microphysical systems
is impossible.
\end{enumerate}

Lately, there have been several exceptions to the widespread
conclusion that the time evolution described by the Hamiltonian must
be time-reversible.  Different people mentioned different reasons why
a microphysical ``arrow of time'' should exist.
\begin{list}
{\roman{vns}}{\usecounter{vns}}
\item Peierls~\cite{Peierls} and his school 
emphasized the importance of the initial
and boundary conditions.  They chose purely outgoing boundary conditions
 for the solutions of the Schr\"{o}dinger
equation.  According to Peierls' ideas, irreversibility is connected with the
choice of the boundary or initial conditions (for the ``wave functions'').

\item T.~D.~Lee~\cite{Lee} explained that the time reverse of a decay
process is impossible or highly improbable to attain due to the phase
of the state vector of the quantum system.  These arguments
immediately extend to scattering experiments.
\begin{quote}
It is easy to prepare two uncorrelated incoming beams that scatter
into strongly correlated outgoing spherical waves as done in a typical
scattering experiment.  It is experimentally hopeless to prepare a
state consisting of two strongly correlated spherical waves (with
fixed relative phase) in such a way that after the scattering of two
uncorrelated plane waves emerge.  The latter would be the time reverse
of the setup for a typical scattering experiment.
\end{quote}

\item Ludwig~\cite{Ludwig} had also noticed that in an experiment with
quantum systems one could not time translate the trigger of a detector
that registers an observable to a time before the preparation
apparatus (e.g.~accelerator) has been turned on.  This 
preparation\-$\longrightarrow$\-registration arrow of time 
for the apparatuses ought to be transcribed into a
semigroup evolution of the state (defined by the preparation
apparatus) relative to the observables (defined by the
detector)\footnote{cf.~Sect.\ \ref{Timesym}, 
footnote (\ref{staobfn}).}.  However, knowing
that in HS the time evolution is represented by the unitary group, he
extrapolated this semigroup to all times $-\infty<t<\infty$.

\item The semigroup generated by the Hamiltonian emerged unexpectedly
in the mathematical derivation of the time evolution for the Gamow
vectors (\ref{gktpevol}) and therewith introduced microphysical
irreversibility into the RHS formulation of quantum
mechanics~\cite{Bohm2}.

\item Prigogine and his school had emphasized for a long
time~\cite{longt} that irreversibility is fundamental.  It therefore
should be intrinsic to the dynamics of the microsystems rather than
being caused by external effects of a quantum reservoir or the
environment (measurement apparatus).  Irreversibility therefore should
be connected with the Hamiltonian (or Liouvillian) and described by a
non-unitary transformation.

\item I.\ Antoniou~\cite{Anton} showed that the formalism of 
subdynamics leads to the RHS and suggested for this non-unitary
transformation the semigroup of the Gamow vectors generated by the
extended Hamiltonian in the RHS.

\item Coming from an entirely different problem, Gell-Mann and Hartle
\cite{Hartl}
introduced an arrow of time in the quantum mechanics of cosmology.  In 
order to avoid inconsistancies for the probabilities of histories, they 
required a time ordering for the projection operators in a history and the
initial states (cf.~(\ref{twokind}) and (\ref{timeo}) of 
Preface).  This arrow of time is identical to the 
preparation$\longrightarrow$registration arrow of time if one applies the
same formula (viz. (\ref{twokind}) Preface) to the probability to register
the observable (projection operator) in the prepared state $\rho$ of a 
laboratory experiment.
\end{list}
All the examples above are different manifestations of a 
fundamental quantum mechanical distinction 
between the past and the future---different expression of causality in 
quantum physics.
The paradigm of this microphysical irreversibility is the time
evolution of resonances.  Since resonances abound, this time-asymmetry
 is prevalent.  HS quantum
theory cannot describe it, however the same mathematical theory of the
rigged Hilbert space that was introduced to justify the Dirac
formalism also describes the irreversible decay of microsystems and
allows for the mathematical transcription of the quantum mechanical
arrow of time.  It describes irreversibility on the microphysical
level.

\section{From a Pre-Hilbert Space to the Rigged Hilbert Space \label{Froma}}

\footnotesize
\begin{quote}
``A role of rigorous mathematics in physical science is to make sense of 
heuristic ideas (i.e. find the `correct setting')---not to assert they are 
nonsense.''

---M. Fisher\footnote{From talk ``What's Mathematical Physics to Physicists?
Some examples from Past, Present and Future'' by M. Fisher, VIIth 
International Congress on Mathematical Physics, Boulder, Colorado, August 
1983.}
\end{quote}
\normalsize

A pre-Hilbert space is a linear space $\Psi$ with a scalar product.
This scalar product is denoted by
\begin{equation}
\ip{\psi}{F}\ \mbox{or by}\ \bk{\psi}{F}.
\end{equation}
The pre-Hilbert space is without any topological
structure\footnote{cf.~footnote (\ref{dictfn}).}.  Such mathematical concepts
as neighborhoods, convergence of infinite sequences, topological
completeness, continuous operators, continuous functionals, dense
subspaces, etc. are not defined.  This pre-Hilbert space $\Psi$ is
what one usually means in physics when one speaks of the Hilbert
space, and this is what one mostly uses, together with the
representations of $\psi$ by square integrable functions and the
calculation of scalar products as Riemann integrals.  The full
mathematical structure of the Hilbert space \bohmHS\ is much more
complicated, and the proof of statements like the deviation from the
exponential decay law in Sect.\ \ref{Conseq}.1 
and the vanishing of the
transition probability in Sect.\ \ref{Conseq}.2
 require the precise mathematical
definition of \bohmHS.

The linear scalar product space $\Psi$ can be endowed with various
topologies, which means that various definitions for the convergence
of infinite sequences can be given.  We denote these topologies by
$\tau_{\bohmHS}$ for the HS, for example, and label the topological
notions with it, such as $\tau_{\bohmHS}$-convergence.  A space is
completed by adjoining to it the limit elements of all Cauchy
sequences not already contained in the space.

The Hilbert space \bohmHS\ is the completion of the pre-Hilbert space
$\Psi$ with respect to the convergence defined by the norm
$||\phi||=\sqrt{\ip{\phi}{\phi}}$.  If one represents the space $\Psi$
by a space of functions and the scalar product by the usual integral,
then one cannot complete the space $\Psi$ into a Hilbert space \bohmHS\
unless the integral is a Lebesgue integral (as opposed to the more
frequently used Riemann integral).  In this case, each element
$\phi\in\bohmHS$ is represented by a {\em class} of Lebesgue
square-integrable functions which differ on a set of Lebesgue measure
zero (e.g.~all rational numbers), not by one wave function.  This is
the ``realization'' of \bohmHS\ by $L^2({\Bbb R})$.

The RHS is the same linear space $\Psi$ completed with respect to
three different topologies: one stronger, one the same, and one weaker
than the topology defined by the Hilbert space norm.  The stronger
topology $\tau_{\Phi}$, e.g.~a topology defined by a countable number
of norms, leads by completion to the space $\Phi$, which,as a
consequence of its stronger topology, has the property
$\Phi\subset\bohmHS$. The topological dual to $\Phi$, i.e. the space of
{\it continuous} anti-linear functionals on $\Phi$, is denoted by
$\Phi^{\times}$ and since its topology is weaker than the Hilbert
space topology we have $\bohmHS\subset\Phi^{\times}$.  The antilinear
$\tau_{\Phi}$-continuous functionals or functions $F(\phi)$ over the
space of $\phi\in\Phi$ are denoted by $F(\phi)=\bk{\phi}{F}$.  This
bra-ket is an extension of the scalar product $\ip{\phi}{f},\ f\in\bohmHS$
to those elements $F\in\Phix$ which are not elements of \bohmHS.  By
completing $\Psi$ with respect to these three topologies, one obtains
a triplet of spaces.  This is the Gelfand triplet or rigged Hilbert
space~\cite{Gelfand}.
\begin{equation}
\Phi\subset\bohmHS=\bohmHS^{\times}\subset\Phi^{\times},\label{RHStrip}
\end{equation}
\begin{eqnarray}
\mbox{with elements ``bra'' and ``ket''} &
\br{\phi}\in\Phi\;&\kt{F}\in\Phi^{\times}\nonumber\\ \mbox{or ``ket''
and ``bra''} & \kt{\phi}\in\Phi & \br{F}\in\Phi^{\times}
\end{eqnarray}

The vectors $\phi\in\Phi$, either as a ket \kt{\phi} or bra \br{\phi},
represent physical quantities connected with the experimental
apparatuses.  For example, in a scattering experiment the preparation
apparatus (accelerator) defines the state $\phi$ and the observable
\kb{\psi}{\psi} is defined by the registration apparatus (detector),
and both $\phi$ and $\psi$ are elements of $\Phi$.  The vectors \kt{F}
or $\br{F}\in\Phi^{\times}$ represent quantities connected with the
microphysical system, e.g.~scattering states \kt{E} or decaying states
\gktp.

The general observable is now represented by a continuous (also bounded)
 operator $A$ in
$\Phi$ (but in general by an unbounded operator $\overline{A}$ or
$A^{\dag}$ in \bohmHS) and one now has a triplet of operators
corresponding the the triplet of spaces (\ref{RHStrip})
\begin{equation}
A^{\dag}|_{\Phi}\subset A^{\dag} \subset A^{\times}.
\end{equation}
The operator $A^{\dag}$ is the Hilbert space adjoint of $A$ (if $A$ is
essentially self-adjoint, then $A^{\dag}=\overline{A}$, where
$\overline{A}$ denotes the closure of the operator $A$ in \bohmHS).  The
operator $A^{\dag}|_{\Phi}$ is its restriction to the space $\Phi$,
and the operator $A^{\times}$ in $\Phi^{\times}$ is the conjugate
operator of $A$ defined by
\begin{equation}
\bk{A\phi}{F}=\bk{\phi}{A^{\times}F}\ \mbox{for all}\ \phi\in\Phi\
\mbox{and all}\ \kt{F}\in\Phi^{\times}.\label{defAx}
\end{equation}
By this definition, \Ax\ is the extension of the operator $A^{\dag}$
to the space \Phix\ and not the extension of the operator $A$, which
is more often used in mathematics.  The operator \Ax\ is {\em only
defined} for an operator $A$ which is {\em continuous} (and bounded) in
$\Phi$, so \Ax\ is always a continuous operator in \Phix.  In quantum
mechanics it is empirically impossible to restrict oneself to
continuous (and therefore bounded) operators $\overline{A}$ in \bohmHS\
(e.g.~already the observables in $[P,Q]=-i1$ cannot be represented by
$\tau_{\bohmHS}$-continuous operators in \bohmHS, but they can be represented
by $\tau_{\Phi}$-continuous operators in a space $\Phi$ with stronger
topology).  But one can restrict oneself to algebras of observables
\{$A, B,\ldots$\} described by continuous operators in $\Phi$ if the
topology of $\Phi$ is suitably chosen.  Then \{\Ax,
$B^{\times},\ldots$\} are defined and continuous in \Phix.  If $A$ in
(\ref{defAx}) is not self-adjoint then \APhi\ need not be a continuous
operator in $\Phi$ even if $A$ is, but one can still define the
conjugate \Ax\ using (\ref{defAx}), which is a continuous operator in
\Phix.

A generalized eigenvector of an operator $A$ is defined to be that
$F\in\Phix$ which fulfills
\begin{equation}
\bk{A\phi}{F}=\bk{\phi}{\Ax F}=\omega\bk{\phi}{F}\ \mbox{for all}\
\phi\in\Phi,\label{defeval}
\end{equation}
where the complex number $\omega$ is called the generalized
eigenvalue.  This is also written as
\begin{equation}
\Ax\kt{F}=\omega\kt{F}.
\end{equation}
For an essentially self-adjoint operator ($A^{\dag}=\overline{A}=$
closure of $A$) this is often also written as it is in Dirac's
formalism, $A\kt{F}=\omega\kt{F}$, especially if one avoids mentioning
the topological structure and works with just a linear scalar product
space $\Psi$.  The precise meaning of eigenvalue equations like
(\ref{evect}),(\ref{gktpevect}) and (\ref{gktpevol}) is given by
(\ref{defeval}).

The generalized eigenvalue $\omega$ in (\ref{defeval}) may belong to
the continuous spectrum of the Hilbert space operator
$A^{\dag}$%
\newcommand{\Adag}{\ensuremath{A^{\dag}}}%
\newcommand{\tphi}{\ensuremath{\tau_{\Phi}}},
as is usually assumed for Dirac kets, but it need not belong to the
Hilbert space spectrum of \Adag, and for a self-adjoint operator
\Adag\ (i.e. if $A$ is essentially self-adjoint) $\omega$ need not
even be real.  If $\kt{F}\in\bohmHS$, then (\ref{defeval}) is identical to
the definition of an ordinary eigenvector of $\Adag$ in $\bohmHS$ with
discrete eigenvalue $\omega$.

The possible set of generalized eigenvalues of an operator $A$ is
determined by the choice of the space \Phix, or equivalently, by the
choice of $\Phi$ (i.e. by the topology $\tau_{\Phi}$ with which we
choose to equip the linear space $\Psi$).  One can choose the topology
$\tphi$ with respect to which one completes the pre-Hilbert space
$\Psi$ unwisely so that $\Phix$ contains all kinds of things.  Others
have suggested restricting $\tphi$ such that the set of generalized
eigenvalues defined by (\ref{defeval}) is identical with the Hilbert
space spectrum of \Adag~\cite{Napi}.  This would reproduce the Dirac
formalism to the extent that it has been used in the past, because the
Dirac kets are always chosen such that they are connected with the
(absolutely) continuous spectrum of an operator $A^{\dag}=\bar{A}$.
But this would not allow for Gamow kets like those in
(\ref{gktpevect}), and no new physics would be incorporated in such a
``tight rigging''.

We invoke the principle that the space $\Phi$ (and its dual \Phix)
should be chosen by physical arguments.  For one ``kind'' of quantum
physical system (where the term ``kind'' is left to the intuitive
interpretation based on experimental experience) one takes as a
mathematical image one particular space $\Phi$.  For instance, one
could define the topology for $\Phi$ such that the observables of the
physical system under consideration are continuous operators in
$\Phi$~\cite{Roberts,B-n-G}.  
We shall discuss a scattering experiment and
additional physical arguments related to causality and initial
conditions which will be used to specify $\Phi$ further.  Causality
and the choice of initial conditions are old principles of classical
physics which have not been fully utilized in standard quantum
mechanics.

\section{Time-Symmetric Equations and Time-Asymmetric Boundary Conditions
\label{Timesym}}

\footnotesize
\begin{quote}
``The miracle of the appropriateness of 
the language of mathematics for the formulation of the laws of physics is a
wonderful gift which we neither understand nor deserve.''

---E.P.~Wigner\footnote{\label{wigfoot} E.P.~Wigner, Symmetries and 
Reflections, (Ox Bow Press, Woodbridge, Connecticut, 1979), p.~237.}
\end{quote}
\normalsize

Theoretical predictions are based on two prerequisites~\cite{Wijew}:
the {\it laws of nature} and the {\it initial (or boundary)
conditions}.  The laws of nature provide the observables, e.g.~the
algebra of operators derived from a space-time symmetry group or the
Hamiltonian $H$ and the dynamical equations, which are the same in
both the HS and RHS formulations and given by the Schr\"{o}dinger
equation or the von Neumann equation:
\begin{equation}
i\hbar\frac{d \phi(t)}{d t}=H\phi(t)\;\;\ \mbox{or}\;\;\ i\hbar\frac{d
W(t)}{d t}=[H, W(t)].
\end{equation}

The laws of nature are not subject to human influences; they are given
once and forever.  The initial and boundary conditions leave some
freedom of choice but are limited by the achievability---in principle
and in practice---of building experimental apparatuses and by
causality~\cite{Peierls}.

In the standard HS formulation of quantum mechanics, one chooses the
initial and boundary conditions for the \kt{\phi} and the \kt{F} such
that both are always elements in \bohmHS.  In this way one does not take
into consideration the variety of possible choices and limitations for
the initial and boundary conditions due to causality.  In the RHS
formulation, the initial condition can be chosen more specific to the
particular problem under consideration.  We shall use these additional
conditions to define the space $\Phi$, or equivalently the RHS
$\Phi\subset\bohmHS\subset\Phix$, for a quantum scattering system.

The best known example of a RHS is the space in which $\Phi$ is
realized by the Schwartz space of ``well-behaved'' functions,
i.e. functions that have derivatives which are all continuous, smooth
and rapidly decreasing, \bohmHS\ is the space of Lebesgue square
integrable functions, and \Phix\ is the space of tempered
distributions, ${\cal S}\subset
L^2({\Bbb R})\subset{\cal S}^{\times}$.  This is the RHS suitable
for the quantum oscillator whose algebra of observables is represented
by $\tau_{\Phi}$-continuous operators.

In quantum theory, if one distinguishes between preparations and
re\-gis\-tra\-tions\footnote{\label{staobfn}
The one feature on which most discussions of
the foundations of quantum mechanics agree is the dichotomy of
``state'' and ``observable''.  If one interprets quantum theory
objectively from the outside in terms of classical physics, as in
\cite{Ludwig}, then the state is defined by the preparation apparatus
(e.g.~accelerator) and the observable is defined by the registration
apparatus (detector).  Though distinct by their definition, the HS
formulation blurs this differentiation between states and observables
by not specifying which elements of the HS are allowed as ``states''
and which as ``observables''.}, one can further specify the RHS, and
one is led to a pair of RHS's: one representing the preparations and
the other representing the registrations.  As an example we will
discuss a scattering experiment.  The scattering experiment can be
subdivided into a preparation stage and a registration stage.  The
in-state $\phi^{+}$ that evolves from the prepared in-state
$\phi^{\mathrm{in}}$ outside the interaction region is determined by
the preparation apparatus (the accelerator).  The ``out-state''
$\psi^{-}$, detected as the ``out-state'' $\psi^{\mathrm{out}}$
outside the interaction region, is determined by the registration
apparatus (the detector).  According to the physical interpretation of
the RHS formulation, real physical entities connected with the
experimental apparatuses, e.g.~the ensemble \kb{\phi}{\phi} describing
the preparation apparatus or the observable \kb{\psi}{\psi} describing
the registration apparatus are described by the {\em well-behaved}
vectors $\phi, \psi\in\Phi$.  However, states are different from
observables and should be described by a different set of vectors.  We
denote the space of state vectors $\phi^{+}$ by $\Phi_{-}$ and the
space of observable vectors $\psi^{-}$ by $\Phi_{+}$, where
$\Phi=\Phi_{-}+\Phi_{+}$ and $\Phi_{-}\cap\Phi_{+}\neq
0$.%
We will call the elements of \mdo{\Phi} and \pdo{\Phi} {\it very
well-behaved} vectors from below and above, respectively, where below
and above refer to the second sheet of the 
energy surface of the analytically continued
$S$-matrix.  In place of the single rigged Hilbert space we therefore
have a pair of rigged Hilbert spaces:
\begin{eqnarray}
\pup{\phi} & \in & \mdo{\Phi}\subset\bohmHS\subset\tup{\mdo{\Phi}}\ \;\
\mbox{for ensembles or prepared in-states},\label{intrip}\\ \mup{\psi}
& \in & \pdo{\Phi}\subset\bohmHS\subset\tup{\pdo{\Phi}}\ \;\ \mbox{for
observables or registered ``out-states''}.\label{outtrip}
\end{eqnarray}
Here the Hilbert space \bohmHS\ is the same for both triplets, and since we
ignore the discrete eigenvalues here it is $\bohmHS^{\mathrm{a.c.}}$ of the 
continuous spectrum.

Mathematically, these spaces can be defined by their realizations as
function space as in the case of the Schwartz space triplet
${\cal S}\subset L^2({\Bbb R})\subset{\cal S}^{\times}$ above.
The spaces \pdo{\Phi} and \mdo{\Phi} are defined as the spaces that
are realized by the well-behaved Hardy class functions~\cite{Phy43} in
the upper half plane,
${\cal S}\cap\pdo{\bohmHS}^{2}|_{\pup{{\Bbb R}}}$, and in the lower
half plane, ${\cal S}\cap\mdo{\bohmHS}^{2}|_{\pup{{\Bbb R}}}$,
respectively:
\begin{eqnarray}
\pup{\phi}\in\mdo{\Phi} & \mbox{if and only if} &
\bk{\pup{}E}{\pup{\phi}}\in{\cal S}\cap\mdo{\bohmHS}^{2}%
|_{\pup{{\Bbb R}}}\label{inspace}\\
\mup{\psi}\in\pdo{\Phi} & \mbox{if and only if} &
\bk{\mup{}E}{\mup{\psi}}\in{\cal S}\cap\pdo{\bohmHS}^{2}%
|_{\pup{{\Bbb R}}}\label{outspace}
\end{eqnarray} 
The notation $|_{{\Bbb R}^+}$ mean restriction to the positive real
line, i.e. the physical values of energy.
If \bohmHS\ in (\ref{intrip}) and (\ref{outtrip}) denotes the Hilbert
space realized as the space of Lebesgue square integrable functions,
$L^{2}[0, \infty)= L^2({\Bbb R}^+)$, then one can show~\cite{Gad2}
that
\begin{equation}
{\cal S}\cap\pmdo{\bohmHS}^2|_{{\Bbb R}^+}\subset
L^2({\Bbb R}^+)\subset\tup{\left({\cal S}\cap\pmdo{\bohmHS}^2%
|_{{\Bbb R}^+}\right)}\label{Hardys}
\end{equation}
are two rigged Hilbert spaces of functions.  The two 
RHS's of the in-states 
$\{\phi^+\}$ (\ref{intrip}) and of the
out-states $\{\psi^-\}$ (\ref{outtrip}) are mathematically defined as those 
RHS's whose realizations are the two RHS's of ${\cal S}\cap\mdo{\bohmHS}^{2}
|_{\pup{{\Bbb R}}}$ and ${\cal S}\cap\pdo{\bohmHS}^{2}
|_{\pup{{\Bbb R}}}$, respectively.

The opposite sub- and superscripts for vectors and spaces comes from the
opposite nomenclature in physics (scattering theory for $\phi^+,
\psi^-$) and mathematics (theory of Hardy class functions for 
$\Phi_-\doteq{\cal S}\cap\mdo{\bohmHS}^{2}
|_{\pup{{\Bbb R}}}$ and 
$\Phi_+\doteq{\cal S}\cap\bohmHS_+^{2}
|_{\pup{{\Bbb R}}}$) which had been developed independently.  Except for the
nomenclature the spaces are the ``same'', i.e. $\{\phi^+\}=\Phi_-$ and
$\{\psi^-\}=\Phi_+$.  This coincidence is another example of 
``The Unreasonable Effectiveness of Mathematics in Natural 
Sciences''\footnote{cf.~footnote (\ref{wigfoot}).}.

It is of course not obvious at all that the physical spaces
(\ref{intrip}) and (\ref{outtrip}), the mathematical images of the
prepared states and the detected observables respectively, should have
anything to do with the Hardy class spaces (\ref{Hardys}).  Originally
the Hardy class property was introduced in order to derive
(\ref{gktpevect}) as a generalized eigenvalue equation and obtain a
Breit-Wigner energy distribution for the vector associated with the
resonance pole of the $S$-matrix~\cite{Bohm2,Baum2}.  But the Hardy class
property of the in-states $\phi^+\in\Phi_-$ and $\psi^-\in\Phi_+$ is
much more generally valid and can be obtained as a consequence of the
Paley-Wiener theorem~\cite{Phy43} from a mathematical formulation of
causality~\cite{Bohm3}.

The condition of causality can be formulated without any reference to
the mathematical theory of quantum mechanics.  Therefore this
formulation of causality does not depend on the choice of the HS or
RHS formulation.  It is phrased in terms of only the intuitive
properties of the macroscopic preparation and registration
apparatuses.  We call this theory-independent statement of causality
the preparation$\longrightarrow$registration arrow of 
time\footnote{It is curious that the theory-independent version 
(Version 1) of this arrow of time had been
known for some time~\cite{Ludwig}, 
but was then extrapolated ``into the past'' in
order to obtain the unitary one-parameter group $U(t)=e^{i \bar{H} t}$
for the time translation of the observables in the HS.}:

Let $t=t_0 = 0$ denote the point in time before which the preparation
is complete and after which the registration begins.  Then:
\begin{enumerate}
\item Time translation of the registration apparatus relative to the
preparation apparatus makes sense only by an amount $t\geq 0$.
\end{enumerate}
A second version of this statement uses quantum theoretical notions;
we call it the quantum mechanical arrow of time.
\begin{enumerate}\setcounter{enumi}{1}
\item (Heisenberg picture) An observable
\kb{\mup{\psi}(t)}{\mup{\psi}(t)} can be measured in a state
$\pup{\phi}(=\pup{\phi}(0))$ only after the state has been prepared,
i.e., for $t\geq 0$.
\end{enumerate}
or
\begin{enumerate}\setcounter{enumi}{1}
\item (Schr\"{o}dinger picture) A state $\pup{\phi}(t)$ must be
prepared before an observable
$\kb{\mup{\psi}}{\mup{\psi}}(=\kb{\mup{\psi}(0)}{\mup{\psi}(0)})$ can
be measured in that state, i.e., $\pup{\phi}$ must be prepared during
$t\leq 0$.
\end{enumerate}

This has implications for the most fundamental quantities in quantum physics,
the expectation values.  These probabilities ${\cal P}(t)$ are observed
experimentally as a ratio of (large) numbers (detector counts) 
$N(t)/N={\cal P}(t)$ and calculated theoretically as
\begin{subequations}
\begin{eqnarray}
{\cal P}(t) & = &\mathrm{Tr}(\Lambda(t) W)=\mathrm{Tr}(\Lambda W(t)),\ 
\mbox{where}\label{latergold}\\
\Lambda(t) & = & e^{iHt}\Lambda e^{-iHt}\ \ \mbox{and}\ \ 
W(t)=e^{-iHt}We^{iHt}.\label{latergoldb}
\end{eqnarray}
\end{subequations}
where $\Lambda$ denotes the observable and $W$ the state.

In many cases, such as for stationary states and
time-independent observables, the right hand side of (\ref{latergold})
can be calculated for any value of $t$, i.e. the equation is valid
for $-\infty<t<\infty$.  The operators act in the space $\bohmHS^{\mathrm{disc}}$
of the discrete spectrum and all calculations are the same in the RHS 
formulation and the HS formulation.  The physics of stationary states is 
not affected by the arrow of time.

This does not hold generally.  In a scattering experiment, for example, 
the state has to be prepared.
If $t=t_0=0$ is the time at which the preparation of the state 
$W=\kb{\phi^+}{\phi^+}$ is complete and the registration of 
$\Lambda(t)\equiv\kb{\psi^-(t)}{\psi^-(t)}$ can begin, then
\begin{equation}\label{detprob}
\frac{N(t)}{N}={\cal P}(t)=\mathrm{Tr}(\Lambda W(t))=
\mathrm{Tr}(\Lambda(t) W)=
\norm{\bk{\psi^-(t)}{\phi^+}},\ \mbox{for}\ t\geq 0\ \mbox{only}.
\end{equation}
The left hand side, $N(t)/N$ is observed only
for $t\geq 0$ and assumed to be zero for $t<0$ (detector 
clicks before $t=0$ are discounted as noise):
\begin{equation}\label{nodet}
\frac{N(t)}{N} = {\cal P}(t) = 0,\ \mbox{for}\ t<0.
\end{equation}
  In the HS, $W(t)$ of
(\ref{latergoldb}) is in all cases 
calculated for positive as well as negative $t$.
The RHS formulation allows for $W(t)$'s
which can only be {\em calculated} for $t\geq 0$, e.g.~the Gamow states 
$W(t)=\kb{\psi^G(t)}{\psi^G(t)}$ of Sect. \ref{Gamowv}.
If the time scale of the formation of $W(t)$ around $t_0=0$ is orders
of magnitude smaller that the time scale of decay it is always more practical
to work with (\ref{nodet}).

The validity of the preparation$\longrightarrow$registration arrow
(Version 1) is obvious.  With the association between the experimental
and the theoretical quantities of Table I, Version 2 (Heisenberg
picture) should be an obvious consequence of Version 1 if for all
times $t$ these associations are upheld.

In the HS formulation this is not possible, let $t=0$ be the point in
time before which the preparation is completed and after which the
registration begins.  Because in the HS formulation, the time
translation operator $U^{\dag}(t)$ for the state $\phi$ (and $U(t)$
for the observable $\psi$) is defined for $-\infty<t<+\infty$, one can
calculate
$\kb{\psi(\tau)}{\psi(\tau)}=U(\tau)\kb{\psi}{\psi}U^{\dag}(\tau)$ for
{\it any} negative value of $\tau$, and its expectation value in
every state $\phi\in\bohmHS$ is in general not zero for $\tau<0$.  Thus,
in the HS formulation it is possible to calculate a non-zero
expectation value for an observable at times $\tau < 0$, which one
should not observe according to Version~1.

In the RHS formulation the quantum mechanical arrow of time can be
implemented in the following way~\cite{Bohm3}: the prepared state vectors
$\phi^+\in\Phi_-$ representing the preparation apparatus should have a
smooth energy wave function
$\pbk{E}{\phi}\in{\cal S}|_{{\Bbb R}^+}$ whose Fourier transform
${\cal F}(\tau)\in{\cal S}$ should be zero for $\tau > 0$,
because $\tau = 0$ is the time after which there is no preparation.
The vector $\psi^-\in\Phi_+$ describing the observable defined by the
registration apparatus should have a smooth energy wave function
$\mbk{E}{\psi}\in{\cal S}|_{{\Bbb R}^+}$ whose Fourier transform
${\cal G}(\tau)\in{\cal S}$ is zero for $\tau < 0$ because $\tau
= 0$ is the time until the registration apparatus remains turned off.
From this property of the Fourier transforms it follows by the use of
the Paley-Wiener Theorem that the wave functions must have the
properties
\[
\pbk{E}{\phi}\in\bohmHS_-^2\;\;\mbox{and}\;\;\mbk{E}{\psi}\in\bohmHS_+^2
\]
This means they have the properties (\ref{inspace}) and
(\ref{outspace}) which leads to the two different spaces \mdo{\Phi}
and \pdo{\Phi} (\ref{intrip}) and (\ref{outtrip}) for the states and
observables, respectively.  The use of different mathematical spaces
for states (in-states) and observables (so-called ``out-states'') is
one of the new features of the RHS formulation of quantum mechanics.

Neither $\Phi$ nor \bohmHS\ can describe single microsystems as mentioned
above in Sect.\ \ref{Hilbert}.3.  
However, it is intuitively attractive to imagine
that the preparation apparatus acts on the registration apparatus by
the action of single physical entities, the microphysical systems.
The energy distribution for a microphysical system does not have to be
a ``well-behaved'' function of the physical values of energy $E$. For
the entities connected with microphysical systems the RHS formulation
provides the elements of \Phix, \tup{\pdo{\Phi}} and \tup{\mdo{\Phi}}.
In particular, Dirac's scattering states \kt{\mom^{\pm}} are elements
of $\tup{\Phi_{\mp}}$, Gamow's decaying states $\psi^G=\mkt{E_R - i
\Gamma/2}$ are elements of $\tup{\Phi_+}$ and (since resonance poles
come in pairs at the complex energies ($E_R \mp i \Gamma/2$)) the
exponentially growing Gamow vectors \pkt{E_R + i \Gamma/2} are
elements of $\tup{\Phi_-}$.

\section{\label{sechalf}
Consequences of the RHS formulation of Quantum Mechanics}

\subsection{\label{Semigr} Semigroup Time Evolution from Time
Asymmetry and Some Remarks on Time Reversal Transformations}

\footnotesize
\begin{quote}
``When in the 18th century Euler discovered those formulas which till today
delight the mathematical phantasy, he seriously stated that his pencil was more
clever than himself.  This impression that mathematical structures 
can include a kind of self-determination concerns me at this time$\ldots$.
Mathematics and Philosophy attack the world's problems in different ways.
Only by their complementary action do they give the right direction.''

---E. K\"{a}hler\footnote{Preprint (1996), translated by the authors.}
\end{quote}
\normalsize

Once the spaces (\ref{intrip}) and
(\ref{outtrip}),
\begin{eqnarray*}
&\mdo{\Phi}\subset\bohmHS\subset\tup{\mdo{\Phi}}&\ \;\
\mbox{for ensembles or prepared in-states---}\{\phi^+\}\ \mbox{and}\\
&\pdo{\Phi}\subset\bohmHS\subset\tup{\pdo{\Phi}}&\ \;\ \mbox{for
observables or registered ``out-states''---}\{\psi^-\},
\end{eqnarray*}
are chosen, which we did in Sect.~\ref{Timesym} on the basis of some causality
arguments, then it is easy
to see mathematically that the extension of the time
evolution operator $U^{\dag}(t)=e^{-iHt}$ in $\bohmHS$ to an
operator $\tup{\pdo{U}}(t)$ in $\Phix_+$---the conjugate of the operator 
$U(t)$ in $\Phi_+$ as defined by (\ref{defAx})---is only a 
semigroup\footnote{For a semigroup, the inverse $(\tup{U}(t))^{-1}$ does
not have to exist.}.

Historically, this was not the situation because ingrained in our 
thinking was the notion that the time evolution operator
$U(t)=e^{iHt}$ in quantum mechanics---and the representation of any 
continuous symmetry transformation---are unitary (reversible) group operators.
Philosophizing alone would not be enough to take a semigroup instead.  
To arrive at the semigroup, we start
from the empirically desirable properties (\ref{gktpevect}) and 
(\ref{gktpevol}) of Gamow resonance states and let mathematics determine the
path.
Then the result will still be surprising, but now acceptable.

We use the definition of a resonance as a (pair of) first order poles in 
the analytically continued $S$-matrix, $S_{bb'}(E)=\mate{b}{|S(E)|}{b'}$.
In the representation of the $S$-matrix 
\begin{equation}\label{diverge}
\ip{\psi^{out}}{S\phi^{in}}=
\ip{\psi^-}{\phi^+}=\sum_{bb'}\int_{\mathrm{spectrum\ of}\ H}
\bk{\psi^-}{b,E^-}\mate{b}{|S(E)|}{b'}\bk{{}^+b',E}{\phi^+},
\end{equation}
we deform the contour of integration from the cut 
$\{0\leq E <\infty\}$ (the spectrum of $H$) into the second sheet 
of the lower complex plane.  The Dirac kets $\kt{E^-}=\kt{b,E^-}$ (where $b$ 
is the degeneracy quantum numbers) with 
$E\in\{\mbox{continuous (real) spectrum}\}$ are therewith continued to 
complex values. At the position of the
resonance pole $E=z_R=E_R-i\Gamma/2$, they become---using Cauchy's 
formula---the Gamow kets $\kt{z_R^-}$.  This is 
standard scattering theory except that we want to be careful about the 
$\bk{\psi^-}{E^-}$ and the $\bk{{}^+E}{\phi^+}$ and do the analytic 
continuation only for those $\psi^-$ and $\phi^+$ for which it is 
mathematically allowed.  This means that $\bk{\psi^-}{E^-}=\bk{E^-}{\psi^-}^*$ 
and $\bk{{}^+E}{\phi^+}$ must not only be elements of ${\cal S}$, but 
also must be the boundary value of an analytic function.  We repeat the same
process for $\bk{H\psi^-}{E^-}=\mate{\psi^-}{\tup{H}}{E^-}=E\bk{\psi^-}{E^-}$.
In order that (\ref{gktpevect}) holds as a general eigenvalue equation,
\begin{equation}\label{kif}
\mate{\psi^-}{\tup{H}}{z_R^-}=z_R\bk{\psi^-}{z_R^-},
\end{equation}
the $\bk{\psi^-}{E^-}$ must further be required to be of Hardy class,
$\bk{\psi^-}{E^-}\in({\cal S}\cap\bohmHS_-^2)$, in other words, $\psi^-$ 
must be required to be a Hardy class vector, $\psi^-\in\Phi_+$.
Similarly we must require $\phi^+\in\Phi_-$.

Now, in order to calculate (\ref{gktpevol}) in its mathematically precise 
form as a generalized eigenvector equation, one repeats the same process for 
$\psi^-$ in (\ref{diverge}) replaced by $\psi^-(t)=e^{iHt}\psi^-$---the 
observable $\kb{\psi^-}{\psi^-}$ translated by time $t$.  On the right hand
side of (\ref{diverge}) one then obtains 
$\bk{\psi^-(t)}{E^-}=\mate{\psi^-}{e^{i\tup{H}t}}{E^-}=
e^{-iEt}\bk{\psi^-}{E^-}$ in place of $\bk{\psi^-}{E^-}$.  The contour 
deformations needed 
 on the right hand side of (\ref{diverge}) are now possible if and only 
if $t\geq 0$, because then and only then is $e^{-iEt}\bk{\psi^-}{E^-}$ a 
Hardy class function.  Then and only then, one 
obtains, in the same way as 
(\ref{kif}) was obtained, the following result\footnote{
cf.~Sect.~XXI.4, 3rd Ed.~of \cite{Bohm}, and \cite{Bohm2}.}:
\begin{equation}\label{kef}
\mate{\psi^-}{e^{-i\tup{H}t}}{z_R^-}=e^{-iz_Rt}\bk{\psi^-}{z_R^-},
\ \mbox{but only for}\ t \geq 0.
\end{equation}
Thus the Gamow vectors are eigenvectors of the time evolution operator given
by (\ref{kef}) and the time evolution operator on Gamow vectors,
$\tup{\pdo{U}}(t)=e^{-i\tup{H}t}$, is only a semigroup.  After one 
was led to this conclusion and has accepted the semigroup evolution 
for the Gamow vectors, one will have no problems to generalize  
to the whole space $\Phix_+$.

The time evolution operator in the space of microphysical states
\tup{\pdo{\Phi}} (and also the time evolution operator in the space
\tup{\mdo{\Phi}}) is described by a semigroup $\tup{\pdo{U}}(t)$ for
$t\geq 0$ only (analogously in the space \mdo{\Phix} the time
evolution is given by a semigroup $\tup{\mdo{U}}(t)$ for $t\leq 0$).
In the RHS's (\ref{intrip}) and
(\ref{outtrip}) one has the two extensions of the Hilbert space
operator $U^{\dag}(t)$\footnote{Regarding the notation $e^{-iH^\times t}$:
$U(t)$ is the unitary operator in $\bohmHS$ and $e^{iHt}\equiv\sum_{n=0}^\infty
\frac{(it)^n}{n!}H^n$ where the generator of $U(t)$, $H$, is defined on a 
dense subspace of $\bohmHS$, the space  
of differentiable vectors ${\cal D}$, and the 
infinite series converges on the dense subspace ${\cal A}\subset
{\cal D}\subset\bohmHS$ of ``analytic vectors''.  The topological notions of
dense, convergence, differentiable, etc., all refer to the topology in
$\bohmHS$ given by the norm, i.e. $U(t)$ and $U^{\dag}(t)$ are \bohmHS-continuous 
(implying \bohmHS-bounded) operators that are 
\bohmHS-dense, \bohmHS-convergent, etc., in \bohmHS.
The restriction of $U(t)$ to the subspace $\Phi_+$, $U_+(t)=U(t)|_{\Phi_+}$ 
for $t\geq 0$ only, is a $\Phi$-continuous operator; $U^\times_+(t)$ is its
conjugate as defined by (\ref{defAx}).  If one applies $U^\times_+(t)$ to a
generalized eigenvector $F\in\Phix_+$ of $H^\times$ with an 
eigenvalue $\omega$ as defined by (\ref{defeval}), then one can show that
$U^\times_+(t)F=e^{-i\omega t}F$.  For this reason we use the notation 
$U^\times_+(t)=e^{-iH^\times t}$.  The operator 
$H^\times$ is a \Phix-continuous
operator in $\Phix_+$ and one can define $\sum_{n=0}^N\frac{(it)^n}{n!}
H^n\equiv U^N_+$ for all $t\geq 0$.  One would usually want to denote by the
exponential $e^{-iH^\times t}$ the limit with respect to the topology in \Phix
of the sequence $U^N_+$ for $N\rightarrow\infty$.  Whether 
$U^N_+\rightarrow U^\times_+(t)$ for $N\rightarrow\infty$ and/or
whether there is a $\Phi$-dense subspace of vectors $F\in\Phix_+$ on
which $U^N_+F$ (which one would then call $\Phix_+$-analytic vectors) is not
known to us.}
\begin{eqnarray}
\mbox{the conjugate of}\ U|_{\mdo{\Phi}}:\ U^{\dag}(t)\subset
\tup{\mdo{U}} & = & e^{-i\tup{H}t/\hbar};\ \mbox{for}\ t\leq
0\label{inU}\\ \mbox{the conjugate of}\ U|_{\pdo{\Phi}}:\
U^{\dag}(t)\subset \tup{\pdo{U}} & = & e^{-i\tup{H}t/\hbar};\
\mbox{for}\ t\geq 0\label{outU}
\end{eqnarray}
where $\tup{\pmdo{U}}$ denote the extensions of the unitary operator
$U^{\dag}(t)$ to the spaces \pmdo{\Phix}~\cite{PhysicaA}.  Mathematically,
\tup{\mdo{U}} in \mdo{\Phix} can be defined by (\ref{defAx}) only for
values of parameters $t\leq 0$, since for $t>0$ the operator
$U(t)|_{\mdo{\Phi}}$ is not a continuous operator which maps
\mdo{\Phi} into \mdo{\Phi}.  By the same argument, \tup{\pdo{U}} in
\pdo{\Phix} can only be defined by (\ref{defAx}) 
for values of parameters $t\geq 0$
because for $t<0$, $U(t)|_{\pdo{\Phi}}$ is not a continuous operator
which maps \pdo{\Phi} into \pdo{\Phi} as required by (\ref{defAx}).

These are the mathematical arguments by which the semigroup time
evolution is derived from the mathematical properties of the spaces 
$\Phi_-$ and $\Phi_+$, which in Sect.~\ref{Timesym} had been conjectured 
from the intuitive notion of causality.
However without the straightforward mathematical derivation of (\ref{kef})
from (\ref{kif}) and just on the basis of ``philosophical'' causality 
arguments alone, one would probably not have been willing to come up with 
these mathematical properties of $\Phi_-$ and $\Phi_+$ 
that have such drastic consequences like the
semigroup evolution.
As one can see from the details~\cite{PhysicaA,Bohm} of the above  
derivation, the reverse conclusion is also correct:
The
choice of the Hardy class spaces and therewith the 
preparation$\rightarrow$registration arrow follows if one requires 
the existence of the Gamow vectors with the above properties.

The semigroup time evolution of the Gamow vectors is the mathematical
expression of irreversibility for the microphysical decaying states.
As a consequence of the quantum mechanical irreversibility, one can no
longer calculate for every state $W^G(t)=\kb{\psi^G(t)}{\psi^G(t)}$,
with $\psi^G\in\pdo{\Phix}$, another state
$W^{\mathrm{neg}}(t)\equiv\kb{\psi^G(-t)}{\psi^G(-t)}$.  This is
empirically correct but it also leads immediately to the question
whether and how this irreversibility can be compatible with the
definition of a time reversal operator
$A_{T}$\footnote{In place of $A_T$ one could as well use the 
CPT-operator here with minor modifications.}
\newcommand{\bohmAT}{\ensuremath{A_{T}}}, since the state obtained
from a given state $W(t)$ by \bohmAT\ transformation,
\begin{equation}
W^{T}\equiv\bohmAT^{-1}W(t)\bohmAT,
\end{equation}
is usually identified with the state $W^{\mathrm{neg}}\equiv W(-t)$,
i.e. is assumed to have the property
\begin{eqnarray}
\bohmAT^{-1}W(t)\bohmAT & = & W(-t)\label{minust}\\ (\mbox{or}\
\phi^{T}(\mathbf{x}, t) & = & \phi(\mathbf{x}, -t)\ =\
\phi^{*}(\mathbf{x}, t)\ \mbox{for wave functions}).\nonumber
\end{eqnarray}

The answer to this question is that neither the time reversed state
$W^{T}(t)$ nor the backward time translated state
$W^{\mathrm{neg}}(t)$ is in general physically defined.  For example,
in a typical scattering experiment the ``out-states'' represent highly
correlated spherical waves whereas the prepared in-states are
typically two uncorrelated plane waves (e.g., two colliding
monochromatic beams).  The time reversal of this experiment would
require a preparation apparatus that prepares highly correlated (with
fixed phase relationship) incoming spherical waves that would be
scattered into two uncorrelated plane waves.  An experimental setup
that would accomplish this would have to be so complicated that it is
impossible to build, at least in this world.  Thus, not for every
preparable $W$ can one prepare a state which would be described by its
time reversal transformed $W^{T}=\bohmAT^{-1}W\bohmAT$ (for another example
see \cite{Lee}).

The time reversal operator \bohmAT\ is not defined by its action on
states, but by its relation to the observables.  Examples of these
relations are
\begin{equation}
\begin{array}{lcllcl}
\bohmAT P_{i} \bohmAT^{-1} & = & -P_{i}, & \bohmAT J_{i} \bohmAT^{-1} & = & -J_{i},\\
\bohmAT U_{P} \bohmAT^{-1} & = & \varepsilon_{T}\varepsilon_{I}U_{P}, & \bohmAT H
\bohmAT^{-1} & = & H,\\ \bohmAT H_{0} \bohmAT^{-1} & = & H_{0}, & \bohmAT S \bohmAT^{-1} &
= & S^{\dag}=S^{-1},
\end{array}
\end{equation}
and they follow from the extended projective representations of the
extended Poincar\'{e} group~\cite{Wein}.  The generators $P_{i}, H,
J_{i}$ represent momentum, energy, angular momentum, respectively, the
$S$-operator is a complicated function of the interaction Hamiltonian
$V=H-H_{0}$ and $U_{P}$ is the unitary and hermitian parity operator
normalized to $U_{P}^{2}=1$.  The quantities
\begin{equation}
\varepsilon_{T}=\bohmAT^{2},\ \mbox{and}\
\;\varepsilon_{I}=(U_{P}\bohmAT)^{2}\equiv A_{I}^{2}
\end{equation}
are real phase factors which define the four different extensions of
the restricted space-time symmetry transformations by space inversion
$P=g$, time inversion $T=-g$ and the space-time inversion $I=PT=-1$,
which were derived by Wigner~\cite{Wigner}.  Of the four possible
extensions characterized by the four pairs of phase factor
$(\varepsilon_{T},\varepsilon_{I})=(\pm 1, \pm 1)$ the the only
extensions used in relativistic field theory~\cite{Wein} are those
characterized by
\begin{equation}
\varepsilon_{T}=(-1)^{2j}\;\varepsilon_{I}=(-1)^{2j}\;\mbox{where}\ j\
\mbox{is the spin.}\label{normext}
\end{equation}
With this choice the time reversal operator \bohmAT\ has the following
transformation property:
\begin{equation}
\bohmAT:\ \Phi_{\mp}\ \rightarrow\ \Phi_{\pm};\ \ \;\;
\pdo{\Phi}\ni\mup{\psi}=\bohmAT\pup{\phi},\
\pup{\phi}\in\mdo{\Phi}.\label{transpro}
\end{equation}
This is also the solution suggested by conventional scattering theory
where the {\em in}-states $\pup{\phi}$ are the time reverse of the
so-called {\em out}-states $\mup{\psi}$.  The ``out-states'' $\psi^-$
are actually observables and not states because they are specified by
the detector whereas in-states $\phi^+$ are specified by the
preparation apparatus (accelerator).  In our interpretation, the
transformation property (\ref{transpro}) means that states are
transformed into observables and vice-versa.  This requires the
identification of the set of states with the set of observables (i.e.,
no arrow of time) and to assign to every $W(t)$ a
$W^{T}(t)\equiv\bohmAT^{-1} W(t) \bohmAT$ fulfilling (\ref{minust}).  This
means that in the case (\ref{normext}) one cannot have irreversibility
in the sense described above, which as mentioned in Sect. 3 is in
contradiction to at least some arguments concerning the improbability
to prepare time reversed states (cf.\ remark above and Chap.\ 13.2 in
\cite{Lee}).

Fortunately, there are three other classes of representations derived
by Wigner~\cite{Wigner} and one can choose instead of (\ref{normext})
one of the non-standard extensions for \bohmAT\ which lead to a doubling
of the spaces (time reversal doubling~\cite{Wigner}).  We suggest the
choice:
\begin{equation}
\varepsilon_{T}=-(-1)^{2j},\;\;\
\varepsilon_{I}=-(-1)^{2j}.\label{newext}
\end{equation}
The \bohmAT\ which fulfills (\ref{newext}) can be shown to be compatible
with the microphysical irreversibility of (\ref{inU}) and
(\ref{outU})~\cite{Bohm4}.

\subsection{Gamow Vectors \label{Gamowv}}

\footnotesize
\begin{quote}
``The data suggest that a particle decays if it can and that it is stable only
if there is no state$\ldots$ to which it is allowed to decay.  Stability does
not appear to be a criterion for {\em elementarity}.''

---Frauenfelder and Henly\footnote{p. 91-92 of \cite{Frauen}.}
\end{quote}
\normalsize

The RHS formulation
accounts\newcommand{\ermg}{\ensuremath{E_{R}-i\Gamma/2}} for the Dirac
kets; they are defined as generalized eigenvectors over the spaces
$\Phi_+$, $\Phi_-$, and $\Phi=\Phi_+ + \Phi_-$
 with generalized eigenvalues $E$ of the
Hamiltonian\newcommand{\pmup}[1]{\ensuremath{#1^{\pm}}} $H$:
\begin{equation}
\bk{H\phi}{\pmup{E}}=\mate{\phi}{\tup{H}}{\pmup{E}}=E\bk{\phi}{\pmup{E}}\
\mbox{for all}\ \phi\in\Phi,
\end{equation}
where the $E$ represent the scattering energies (continuous spectrum
of $\bar{H}$).  This is no surprise because the RHS formulation was
devised to provide a mathematical justification for the Dirac
scattering states.

Unforeseen was that the RHS formulation also accounts for the
Gamow kets.  The decaying Gamow kets 
$\psi^{G}=\kt{\mup{\ermg}}\sqrt{2\pi\Gamma}=\mkt{z_R}\sqrt{2\pi\Gamma}$ 
are defined as generalized
eigenvectors over the space \pdo{\Phi} with generalized eigenvalue
(\ermg):
\begin{subequations}\label{actofH}
\begin{equation}
\bk{H\mup{\psi}}{\mup{\ermg}}  \equiv 
\mate{\mup{\psi}}{\tup{H}}{\mup{\ermg}} = 
(\ermg)\bk{\mup{\psi}}{\mup{\ermg}},
\end{equation}
or its complex conjugate:
\begin{equation}
\mate{\psi^G}{H}{\psi^-}  = \bk{\psi^G}{\psi^-}(E_R+i\Gamma/2),\ 
\mbox{for all}\ \mup{\psi}\in\pdo{\Phi}.
\end{equation}
\end{subequations}
These Gamow vectors $\psi^{G}=\kt{\mup{\ermg}}\sqrt{2\pi\Gamma}$ have
the following properties:
\begin{list}
{(\arabic{gv})}{\usecounter{gv}}
\item They are derived as functionals of the resonance pole term at
$z_{R}=\ermg$ in the lower half of the second sheet of the
analytically continued $S$-matrix.

\item They have Breit-Wigner energy distribution
\begin{equation}
\bk{\mup{}E}{\psi^{G}}=i\sqrt{\Gamma/2\pi}\frac{1}{E -
(\ermg)},\;\;\;\ -\infty_{II}<E<\infty,\label{BWd}
\end{equation}
where $E$ is on the second sheet for negative values\footnote{
\label{general} The ``wave function'' $\bk{{}^-E}{z_R^-}$ of the generalized
vector $\kt{z_R^-}$ in (\ref{BWd}) 
is represented by a regular function (even of 
Hardy class $\bohmHS_+^2$) though one would expect it to be a non-regular 
distribution.  The latter is indeed correct because (\ref{BWd}) is 
not the actual wave function.  The wave function is a
function on the continuous spectrum $\{0\leq E \leq\infty\}$ and over
${\cal S}\cap\bohmHS_+^2|_{{\Bbb R}^+}$, i.e.~as an element of 
$({\cal S}\cap\bohmHS_+^2|_{{\Bbb R}^+})^\times$, $\bk{{}^-E}{z_R^-}$ is
 a distribution---not a function (regular distribution).  
However in (\ref{BWd}) we need the whole real axis $\{-\infty<E<\infty\}$ to
represent it as a regular function. Extending $E$ in 
(\ref{BWd}) to the negative real axis does however not mean that 
we have let the 
spectrum of $H$ go to $-\infty$, because the negative values of $E$ in 
(\ref{BWd}) are the result of an analytic continuation into 
the second sheet.  For more detail see p. 504-505 of 
\cite{PhysicaA}, or see \cite{toappear}, where an explicit expression for the 
distribution 
$\bk{{}^-E}{z_R^-}\in({\cal S}\cap\bohmHS_+^2|_{{\Bbb R}^+})^\times$
has been given.}.

\item The time evolution of the Gamow vectors is derived from the pole
term of the $S$-matrix as:
\begin{subequations}\label{decst}
\begin{equation}
\bk{e^{iHt}\mup{\psi}}{\mup{z_{R}}}%
\equiv\mate{\mup{\psi}}{e^{-i\tup{H}t}}{\mup{z_{R}}}%
=e^{-iE_{R}t}e^{-(\Gamma/2)t}\bk{\mup{\psi}}{\mup{z_{R}}}\label{decayst}
\end{equation}
or for the complex conjugate:
\begin{equation}\label{decaystcc}
\mate{\mup{}z_{R}}{e^{iHt}}{\mup{\psi}}=e^{iE_{R}t}%
e^{-(\Gamma/2)t}\bk{\mup{}z_{R}}{\mup{\psi}},\
\mbox{for every}\ \mup{\psi}\in\pdo{\Phi}\ \mbox{and for}\ t\geq 0.
\end{equation}
\end{subequations}
\end{list}

The above properties---except for the semigroup property---are 
historically ascribed to the empirical notion of  
decaying states and resonances.  Equations 
(\ref{actofH}) and (\ref{decst}) 
give the precise mathematical form as a distribution.  Therefore the 
Gamow kets justly deserve the name
resonance state vectors with complex resonance energy $z_{R}=\ermg$.
The property (2) identifies $E_{R}$ as the resonance energy and
$\Gamma$ as the resonance width.  The property (3) shows that
$\tau_{R}=1/\Gamma\,(=\hbar/\Gamma)$ is the lifetime of the decaying
resonance state.

The semigroup evolution (\ref{decst}) expresses the time-asymmetry of the 
Gamow vector and the resonance state which it describes.  It is not a 
condition that we demanded of the \kt{z_R^-}, like we demanded (\ref{detprob})
or some other mathematical formulation of causality, but it can be derived 
from the Hardy class property (\ref{outtrip}) which is some mathematical 
statement of causality\footnote{
It is remarkable that the asymmetry of the 
time evolution of the resonance 
states (\ref{decst})---or more generally of the semigroup (\ref{outU})---is
the same as the preparation$\rightarrow$registration arrow of time 
(\ref{detprob}).  This is established mathematically by the use of the
Hardy class RHS's (\ref{intrip}) and (\ref{outtrip}).  Since equation 
(\ref{detprob}) applied to the quantum theory of cosmology is, according
to (\ref{preprob}), the time-asymmetry of the universe~\cite{Hartl}, the mathematical
idealization given by the RHS formulation of quantum mechanics would
establish a connection between the time-asymmetry of quantum cosmology and
the irreversibility of quantum decay phenomena.}.

In the same way as stationary states are given by bound state
poles of the $S$-matrix and described as eigenvectors 
of $H$ with real (negative) eigenvalues, decaying states are
given by resonance poles and described as generalized eigenvectors of
$H$ with complex eigenvalue.  This puts stable and decaying particles
on the same footing and shows that resonances can be thought of as
autonomous quantum physical entities which are not less fundamental
than stable particles.

Instead of defining the Gamow vectors from the poles of the analytically
continued $S$-matrix, the Gamow vectors can also be defined from the poles
of the extended resolvent of the Hamiltonian~\cite{lastminute}.  
These definitions
are equivalent in many cases in which the $S$-matrix is obtained from the
Hamiltonian.  The $S$-matrix definition can be used in more general settings,
e.g.\ in the relativistic case when one can start from the unitary 
representations of the Poincar\'{e} group in \bohmHS\ and then extend it to
$\tup{\Phi}_\pm$.

One can generalize the Gamow vectors to Gamow-Jordan vectors associated to
higher-order poles of the $S$-matrix~\cite{AntGad} and obtain higher-order
Gamow states~\cite{Bohm5}.  These vectors are also functionals over
Hardy class spaces and have semigroup time evolutions.  They have some 
features which are heuristically associated with higher-order poles of the 
$S$-matrix~\cite{Goldberg} like polynomial time dependence.  But 
non-reducible Gamow states that follow from the higher-order $S$-matrix pole
term have purely exponential time evolution, which is a feature that was not 
expected.  Since so far there is no empirical evidence for states associated
with higher-order poles of the $S$-matrix, we shall not discuss them here any 
further.

\subsection{Generalized Basis Vector Expansions}
The most important consequence of the RHS formulation are the
generalized basis vector expansions.  The expansion (\ref{expans})
that Dirac envisioned has been proven as the nuclear spectral theorem.
This theorem requires the nuclearity of the space $\Phi$ (or a little
less~\cite{Gelfand}), a mathematical subtlety unimportant for physics.
What is important for physics is that (\ref{expans}) holds in the
space $\Phi$; therefore we will call a triplet of spaces
(\ref{RHStrip}) a RHS only if the necessary conditions for the nuclear
spectral theorem are fulfilled.

All basis vector expansions are generalizations of the elementary
basis vector expansion of a vector in ${\Bbb R}^3$,
\begin{equation}
\mathbf{x}=\sum_{i=1}^{3}\hat{\mathbf{e}}_{i}%
(\hat{\mathbf{e}}_{i}\cdot\mathbf{x})%
=\sum_{i=1}^{3}\hat{\mathbf{e}}_{i}\cdot
x_{i},\label{element}
\end{equation}
where $\hat{\mathbf{e}}_{i}$ are chosen to be the physically
distinguished basis vectors.

Earlier generalizations of this are the fundamental theorem of linear
algebra which states that for every self-adjoint operator $H$ in an
$n$-dimensional Euclidean space $\bohmHS_{n}$ there exists a complete
basis system $e_{i}\ldots e_{n}$ in $\bohmHS_{n}$ of eigenvectors:
\begin{eqnarray}
He_{i} & = & E_{i}e_{i},\ e_i\in\bohmHS_n\ (i=1,2,\ldots,n),\ \mbox{such
that}\nonumber\\ f & = & \sum^n_{i=1}e_i\dbk{e_i}{f}\ \;\mbox{for
every}\ f\in\bohmHS_n\label{Euc}
\end{eqnarray}
This theorem generalizes to the infinite dimensional Hilbert space
\bohmHS, but only for self-adjoint operators $H$ which are completely
continuous\footnote{These are also called compact operators and
include Hilbert-Schmidt, nuclear and trace-class operators.}.  For an
arbitrary self-adjoint operator $H$ one cannot find a complete system
of eigenvectors in \bohmHS\ (complete in the sense that every $f\in\bohmHS$ can
be expanded in the form (\ref{Euc})).  Many physically important
operators do not have even a single eigenvectors in \bohmHS.  Because
\norm{\ip{e_{i}}{f}} represents the probability to measure the value
$E_{i}$ of the observable $H$ in the state $f$, one wants to use a
basis system of eigenvectors for the distinguished observables not an
arbitrary basis system, which one can always find by the Schmidt
orthonormalization procedure.  In order to obtain such a basis system
of eigenvectors for an arbitrary observable, one has to go outside the
HS; they are generalized eigenvectors as defined by (\ref{defeval}).

The {\em first} step outside the HS is to use the nuclear spectral
theorem to justify the Dirac basis vector expansion (\ref{expans}) in
terms of well-defined mathematical quantities.  The nuclear spectral
theorem states that for every $\phi\in\Phi$ (not for every $f\in\bohmHS$)
one can find a complete set of eigenvectors in $\Phix$ (not in \bohmHS):
\begin{subequations}\label{fbexpp}
\begin{equation}\label{fbexppa}
\phi=\int^{+\infty}_{0}dE\kt{\pup{E}}\bk{\pup{}E}{\phi} +
\sum_{n}\dkt{E_{n}}\dbk{E_{n}}{\phi}\ \mbox{for every}\ \phi\in\Phi.
\end{equation}
Since $\phi^+\in\Phi_-\subset\Phi$, of course the same expansion holds
for $\phi^+$:
\begin{equation}
\phi^+=\int^{+\infty}_{0}dE\kt{\pup{E}}\bk{\pup{}E}{\phi^+} +
\sum_{n}\dkt{E_{n}}\dbk{E_{n}}{\phi^+}.\label{fbexppb}
\end{equation}
\end{subequations}
In the above expansion, \dkt{E_{n}} are the discrete eigenvectors of
the exact Hamiltonian $H=K+V$, $H\dkt{E_{n}}=E\dkt{E_{n}}$, which
describe bound states.  The generalized eigenvectors (Dirac kets) of
$H$, $\kt{\pup{E}}\in\Phix$, fulfill
$\bk{H\chi}{\pup{E}}=\mate{\chi}{\tup{H}}{\pup{E}}=E\bk{\chi}{\pup{E}}$
for all $\chi\in\Phi$, cf.\ (\ref{defeval}).  The ``coordinates'' of
the vector $\phi$ with respect to the continuous basis \pkt{E}, i.e.,
the set of energy wave functions \pbk{E}{\phi}, form a realization of
the space $\Phi$ by a space of functions (in the same way as the
coordinates $x_i$ form a ``realization'' of the vector $\mathbf{x}$).
We call the vector $\phi\in\Phi$ ``well-behaved'' if
\bk{\pup{}E}{\phi} is a well-behaved function, i.e.,
$\bk{\pup{}E}{\phi}\in{\cal S}$.  The \pkt{E} correspond to the
continuous spectrum of $H$ (the ``scattering states'') and the
integration extends over the continuous spectrum
\footnote{\label{prereq}
The nuclear spectral theorem actually asserts much less than (\ref{fbexpp}) 
but it applies to a much larger set of operators $A$ than the operators
needed in quantum physics:

Let $A$ be {\em any} self-adjoint (or unitary or normal) operator in 
the Hilbert space $\bohmHS$ (the analogous statement holds for a strongly
commuting family of operators $A_1,A_2,\ldots,A_N$) of the rigged Hilbert
space $\Phi\subset\bohmHS\subset\Phix$, and $\Lambda$ its Hilbert space spectrum.  Then
there exists a unique measure $\mu$ on $\Lambda$ such that for $\phi,
\psi\in\Phi$
\begin{equation}\label{prime}
\ip{\psi}{\phi}=\int_\Lambda\:d\mu(\lambda)\bk{\psi}{\lambda}
{\lambda}{\phi}\tag{\ref{fbexpp}$'$}
\end{equation}
where the $\kt{\lambda}\in\Phix$ are the generalized eigenvectors of $A$, i.e.
\begin{equation}\label{primeprime}
\bk{A\phi}{\lambda}\equiv\mate{\phi}{A^\times}{\lambda}=\lambda
\bk{\phi}{\lambda},\ \mbox{for $\mu$-almost every $\lambda$}.
\tag{\ref{fbexpp}$''$}
\end{equation}
The measure $d\mu(\lambda)$ depends upon the operator $A$, and for a general 
operator $\mu$ consists of three parts:
\begin{enumerate}
\item the discrete spectrum, $d\mu(\lambda)=\sum_n \delta(\lambda-
\lambda_n)d\lambda$,
\item the absolutely continuous spectrum, $d\mu(\lambda)=\rho(\lambda)
d\lambda$, and 
\item the singularly continuous spectrum.
\end{enumerate}
The discrete spectrum corresponds to the sum in (\ref{fbexpp}), the 
absolutely continuous spectrum corresponds to the integral in (\ref{fbexpp})
and the singularly continuous spectrum has never occurred for operators
that represent physical observables.

The observables in quantum mechanics are usually derived from representations
of space-time symmetry groups, spectrum generating groups, intrinsic
symmetry groups or other Lie groups.  Therefore in physics the measure
$d\mu$ is the Plancherel measure of compact or non-compact Lie groups and
their compact or non-compact subgroups.  For the compact case, one has only
case (1); for the non-compact Lie groups that have occurred in
physics (e.g. all Abelian groups or all classical groups, semi-direct 
product and more) one has case (2) with $\rho(\lambda)$ being a continuous
function.  In these cases one has the ``normalizations'' of the generalized
eigenvectors:
\[
\bk{\lambda_n}{\lambda_{n'}}=\delta_{nn'}\;\;\bk{\lambda}{\lambda'}
=\rho^{-1}(\lambda)\delta(\lambda - \lambda').
\]
After redefinition of the kets, $\kt{\lambda}\rightarrow\pkt{E}=
\kt{\lambda}\rho^{\frac{1}{2}}(\lambda)$ and $\lambda\rightarrow E$, this
gives (\ref{fbexppa}) with $\bk{E}{E'}=\delta(E-E')$.  The integral in 
(\ref{prime}) is still a Lebesgue integral and (\ref{prime}) holds for
$\phi,\psi\in\bohmHS$ if $\bk{\lambda}{\phi},\bk{\lambda}{\psi}\in
{\cal L}^2(\lambda)$.  However now we can follow the procedure described
in Sect. \ref{Hilbert} and use only
the smooth functions $\bk{\lambda}{\phi},\bk{\lambda}{\psi}\in
{\cal S}_\Lambda$ under the integral in (\ref{prime}). 
 Then for the subspace $\Phi$ whose realization is 
given by the smooth functions ${\cal S}$ we can use in (\ref{prime}) 
Riemann integration with (\ref{primeprime}) holding for every $\lambda$
(rather than $\mu$-almost every $\lambda$).  No 
example of a physical observable is known in which the integrals (\ref{fbexpp})
can not be interpreted as Riemann integrals and the eigenvalue equation is 
valid only for $\mu$-almost all $\lambda$.  We will always presume that the
observables have discrete and/or absolutely continuous spectra.  
Dirac's visionary
tools of quantum mechanics, (\ref{evect}) and (\ref{expans}), have thus
been made rigorous.}
:$0\leq E < \infty$ (the physical
scattering energies).

If the {\em out}-wave functions are more
readily available, one would have chosen the \mkt{E}, defined as
$\pkt{E}=\mkt{E}S(E + i0)$, in place of the $\pkt{E}$ in
(\ref{fbexpp}), or one could have chosen any other vector which
differs from \pkt{E} by an energy-dependent phase factor\footnote{
\label{channelfn} If
the additional quantum numbers (e.g., channel labels) $\eta$ would be
considered, then in place of the phase factor $S(E)=e^{2i\delta(E)}$
one would of course have a unitary matrix, $\pkt{E,\ \eta}=\mkt{E,\
\eta'{}}\mate{\eta'}{S(E)}{\eta}$.}.
\begin{eqnarray}
\phi & = & \int^{+\infty}_{0}dE\mkt{E}S(E + i0)\bk{\pup{}E}{\phi} +
\sum_{n}\dkt{E_{n}}\dbk{E_{n}}{\phi}\nonumber \\ \phi & = &
\int^{+\infty}_{0}dE\mkt{E}\bk{\mup{}E}{\phi} +
\sum_{n}\dkt{E_{n}}\dbk{E_{n}}{\phi}\ \ \mbox{for every}\
\phi\in\Phi\label{fbexpm}
\end{eqnarray}

In a scattering experiment the following $S$-matrix elements are
measured
\begin{eqnarray}
\ip{\psi^{\mathrm{out}}}{\phi^{\mathrm{out}}} & = &
\ip{\psi^{\mathrm{out}}}{S\phi^{\mathrm{in}}} =
\ip{\mup{\psi}}{\pup{\phi}}\nonumber\\ & = &\int_{0}^{+\infty}
\int_{0}^{+\infty}dE\,dE'\,\bk{\psi^{\mathrm{out}}}{E}%
\mate{E}{S}{E'}\bk{E'}{\phi^{\mathrm{in}}}\nonumber\\
& = & \int_{0}^{+\infty}dE\mbk{\psi}{E}S(E + i0)\pbk{E}{\phi}
\end{eqnarray}
In here $\mup{\psi}$ and $\pup{\phi}$ are very well-behaved vectors
with:
\begin{quote}
$\pup{\phi}\in\mdo{\Phi}$ representing the state prepared by the
accelerator, cf.\ (\ref{intrip}) and \\ $\mup{\psi}\in\pdo{\Phi}$
representing the observable registered by the detector, cf.\
(\ref{outtrip}).
\end{quote}
This means that the functions
$\bk{E}{\psi^{\mathrm{out}}}=\mbk{E}{\psi}$ are ``very well-behaved
functions from above'' and $\bk{E}{\phi^{\mathrm{in}}}=\pbk{E}{\phi}$
are ``very well-behaved functions from below'', i.e., equations
(\ref{inspace}) and (\ref{outspace}):
\begin{subequations}
\begin{equation}
\bk{E}{\phi^{\mathrm{in}}} =
\pbk{E}{\phi}\in{\cal S}\cap\mdo{\bohmHS}^{2}|_{\pup{{\Bbb R}}}
\label{inspace2}
\end{equation}
\begin{equation}
\bk{E}{\psi^{\mathrm{out}}} =
\mbk{E}{\psi}\in{\cal S}\cap\pdo{\bohmHS}^{2}|_{\pup{{\Bbb R}}}.
\label{outspace2}
\end{equation}
\end{subequations}
In the scattering experiment, the wave functions \pbk{E}{\phi} (the
components of the vector $\pup{\phi}$ along \pkt{E}) represent the
probability $\int_{\Delta E}dE\norm{\pbk{E}{\phi}}$ that the beam
prepared by the accelerator has an energy in the interval $\Delta E$,
i.e., \norm{\pbk{E}{\phi}} is the energy distribution in the beam
state $\pup{\phi}$ and \norm{\mbk{E}{\psi}} is the energy resolution
of the detector $\mup{\psi}$ (the detector efficiency).  But the
components of $\pup{\phi}$ along \mkt{E}, and of $\mup{\psi}$ along
\pkt{E}, do not just represent apparatus properties, but the
properties of the scattering system.

If we consider for the vector $\phi$ in (\ref{fbexpm}) a very
well-behaved vector $\pup{\phi}\in\mdo{\Phi}\subset\Phi$
\begin{equation}
\pup{\phi}=\int_{0}^{\infty}dE\,\mkt{E}\mpbk{E}{\phi} +
\sum\dkt{E_{n}}\dbk{E_{n}}{\pup{\phi}}\label{fbexpmp}
\end{equation}
then we obtain a Dirac basis vector expansion of $\phi^+$ with respect
to the generalized eigenvectors \mkt{E} of $H$.  Although both
expansions (\ref{fbexppb}) and (\ref{fbexpmp}) use generalized
eigenvectors of $H$, the basis system used in expansion
(\ref{fbexpmp}) differs from that of (\ref{fbexppb}) by a phase
factor\footnote{cf.\ footnote (\ref{channelfn}).}.  Whereas the components
\pbk{E}{\phi} along \pkt{E} contain only information about the
preparation apparatus, the components \mpbk{E}{\phi} are the
interaction wave functions describing also the dynamics (analogous
considerations hold for ``out-state'' (observable) vectors
$\psi^-\in\Phi_+\subset\Phi$).

The {\em second} step on the way outside the HS is the {\em complex
basis vector expansion}.  It holds for ``very well-behaved'' vectors,
i.e., for vectors of a subspace $\mdo{\Phi}\subset\Phi$ only.  For
every $\pup{\phi}\in\mdo{\Phi}$ (and a similar expression holds also
for every $\mup{\psi}\in\pdo{\Phi}$), one obtains the following basis
vector expansion for the case of an $S$-matrix with a finite number of
resonance poles at the positions $z_{R_{i}}=E_{R_{i}} - i
\Gamma_{i}/2$, $i=1,2,\ldots,N$:
\[
\pup{\phi} = \int^{-\infty_{II}}_{0} dE\,\mkt{E}\mpbk{E}{\phi} -
\sum_{i=1}^{N}\mkt{z_{R_{i}}}2\pi\Gamma_{i}\pbk{z_{R_{i}}}{\phi} +
\sum_{n}\dkt{E_{n}}\dbk{E_{n}}{\phi^+}\]
\begin{equation}
\mbox{for}\ \pup{\phi}\in\mdo{\Phi}.\label{cfbexpmp}
\end{equation}
Here
$\mkt{z_{R_{i}}}\sqrt{2\pi\Gamma_{i}}=\psi^{G_{i}}\in\tup{\pdo{\Phi}}$
are Gamow kets (\ref{actofH}) representing decaying states
(\ref{decayst})\footnote{There is a corresponding basis vector
expansion of $\mup{\psi}\in\pdo{\Phi}$; in place of \mkt{z_{R_{i}}},
it contains generalized eigenvectors \pkt{z^{*}_{R_{i}}} with
eigenvalue $z^{*}_{R_{i}}=E_{R_{i}} + i \frac{\Gamma_{i}}{2}$.  These
are the Gamow vectors associated with the $S$-matrix pole at
$z_{R_{i}}^*$.  They have an exponentially growing semigroup evolution
for $-\infty<t\leq 0$.}.  The remarkable feature of the basis vector
expansion (\ref{cfbexpmp}) is that the decaying states \mkt{z_{R_{i}}}
appear on the same footing as the stationary states
\dkt{E_{n}}\footnote{The forms (\ref{fbexpp}) and (\ref{cfbexpmp}) of
the generalized basis vector expansions assume that $H$ is the only
observable to be diagonalized (cyclic operator).  If the complete
system of commuting observables (c.s.c.o.) consists of $(N+1)$
operators $H, B_{(1)}, B_{(2)},\ldots,B_{(N)}\equiv H,B$, then we have
to make the following replacements for the projection operators
\[
\dkb{E_{n}}{E_{n}}\rightarrow\sum_b\dkb{E_{n}, b}{E_{n}, b},
\]
where the sum extends over all values of the degeneracy quantum
numbers $b=b_{(1)},b_{(2)}\ldots b_{(N)}$ of the energy
$E_{n}$.  Similarly, in (\ref{fbexpp}) to (\ref{cfbexpmp}) we have to
make the replacements:
\begin{eqnarray}
\pkb{E}{E} & \rightarrow & \sum_{b}\pkb{E,b}{E,b}\\
&=&\int\:d\mu(b_1,b_2,\ldots)\sum_{b_d,\ldots,b_N}
\pkb{E, b_1\ldots b_N}{E, b_1\ldots b_N}\nonumber\\
\mpkb{z_R}{z_R} & \rightarrow & \sum_b\mpkb{z_R, b}
{z_R,b}\nonumber\\
&=&\int\:d\mu(b_1,b_2,\ldots)\sum_{b_d,\ldots,b_N}
\mpkb{z_R,b_1\ldots b_N}{z_R,b_1\ldots b_N}\nonumber
\end{eqnarray}
where $b_1,b_2,\ldots$ are the continuous and $b_d,\ldots,b_N$ are the
discrete degeneracy quantum numbers.
The operators $B$ could be, e.g., the orbital angular momentum
operators $J_3,\mathbf{J}^2$ if we have a spherically symmetric
(spin-less) scattering system $[H, J_i]=0$; then the quantum numbers
$b=b_1,b_2$ are $b=j,j_3$.  They could be the momentum operators $P_i$ if
$[H,P_1]=0$; then $(E,b_1,b_2)=(p_1,p_2,p_3)$ or $(E,b_1,b_2)=(E,
\theta_p,\varphi_p)$, where $(\theta_p,\varphi_p)$ are the spherical 
coordinates of $\mathbf{p}$ and $d\mu(b_1,b_2)=d\cos\theta_p d\varphi_p.$
The labels $b$ could also be some intrinsic
quantum numbers, like charges or the channel label $\eta$.}.

The last term in the expansions (\ref{fbexpp}), (\ref{fbexpmp}) and
(\ref{cfbexpmp}) will be absent if there are no bound states; we shall
omit this term in the following discussions.  The first term in
(\ref{cfbexpmp}) is the background integral related with the
background phase shifts.  The integration, taken along the negative
real axis in the second sheet (for which the values of
$\mpbk{E}{\phi}=S(E)\pbk{E}{\phi}$ can be calculated from the
experimental values \bk{E}{\phi^{\mathrm{in}}}\ for
positive energies using the van Winter theorem~\cite{vanWinter}),
could be deformed into integration over many other equivalent contours
in the lower half plane of the second sheet, if those would be more
convenient to calculate.

For a $\pup{\phi}\in\mdo{\Phi}\subset\Phi$, both expansions
(\ref{fbexpmp}) and (\ref{cfbexpmp}), which use two different basis
systems, hold.  The expansion (\ref{cfbexpmp}) separates the
individual resonance poles, whereas (\ref{fbexpmp}) has the resonances
contained together with the ``background'' in the energy wave function
\mpbk{E}{\phi}.  The function \norm{\mpbk{E}{\phi}}\ may have a bump
at $E=E_{R}$, but $\mpbk{E}{\phi}\in{\cal S}$ cannot be the
Breit-Wigner distribution (\ref{BWd}) characteristic of a resonance
because the Breit-Wigner distribution is not a well-behaved
function\footnote{Note that the Gamow ket has an ``idealized''
Breit-Wigner energy distribution \bk{{}^-E}{\psi^G} (\ref{BWd}) which
extends from $-\infty_{II}$ to $+\infty_{II}=\infty_I$.  This is a
regular function on ${\Bbb R}$, but $\psi^G$ cannot be realized by a regular
function on the physical energy values ${\Bbb R}_+=\{ E |\ 0\leq E <
\infty_I \}$ since $\psi^G$ is a generalized vector and
correspondingly \bk{{}^-E}{\psi^G} is a distribution over
${\cal S}\cap\bohmHS_+^2|_{{\Bbb R}^+}$, cf. also 
footnote \ref{general}.}.  The complex energy basis
vector expansion (\ref{cfbexpmp}) is the much preferred representation
for investigating resonances.

For the sake of definiteness we shall now assume that there are two
decaying states $R_{1}=S$ and $R_{2}=L$ and no bound states.
According to the expansion (\ref{cfbexpmp}), the pure state (prepared
by the experimental apparatus) has the following representation in
terms of the Gamow vectors
$\psi^{G}_{L}=-\mkt{z_{L}}\sqrt{2\pi\Gamma_L}$%
\newcommand{\pgl}{\ensuremath{\psi^{G}_{L}}},
$\psi_{S}^{G}=-\mkt{z_{S}}\sqrt{2\pi\Gamma_S}$%
\newcommand{\pgs}{\ensuremath{\psi^{G}_{S}}}
and the remaining part which we call
$\phi_{\mathrm{bg}}^{+}$\newcommand{\pbg}{\ensuremath{\phi^{+}_{\mathrm{bg}}}}
(the background):
\begin{equation}
\pup{\phi}=\pgl b_{L} + \pgs b_{S} +
\int_{0}^{-\infty_{II}}dE\,\pkt{E}\pbk{E}{\phi}.\label{lsexp}
\end{equation}
In here $b_{L}$ and $b_{S}$ are some complex numbers that depend on
the normalization of the Gamow vectors \pgl, \pgs\ (and of
\pup{\phi}), and upon some phase convention.  All the vectors in the
generalized basis system expansion are (generalized) eigenvectors of
the exact Hamiltonian, and, in particular, the Gamow vectors \pgl,
\pgs\ are eigenvectors of the exact Hamiltonian with complex
eigenvalue $(E_{L} - i\Gamma_{L}/2)$ and $(E_{S} - i\Gamma_{S}/2)$,
respectively.  If we ignore $\pbg=\int^{-\infty_{II}}_0
dE\,\pkt{E}\pbk{E}{\phi}$ then \pup{\phi} in (\ref{lsexp}) is the
superposition of two eigenvectors of $H$ with complex eigenvalues
$z_{L}$ and $z_{S}$; the Hamiltonian matrix is complex and
diagonalizable.

We now apply the time evolution operator to (\ref{lsexp}).  Since the
\pgl, \pgs\ are elements of \tup{\pdo{\Phi}} we apply the operator
$\tup{\pdo{U}}(t)$ of (\ref{inU}) and obtain:
\begin{equation}
\pup{\phi}(t)\equiv e^{-i\tup{H}t}\pup{\phi} =
e^{-i(E_{L}-i\Gamma_{L}/2)t}\pgl b_{L} +
e^{-i(E_{S}-i\Gamma_{S}/2)t}\pgs b_{S} + \pbg(t);\ t\geq
0.\label{tevolls}
\end{equation}
Since the time evolution semigroup (\ref{inU}) has the restriction
$t\geq 0$, the same restriction must be used for (\ref{tevolls}).  The
time evolved background term is
\begin{equation}
\pbg(t)\equiv\int_{0}^{-\infty_{II}}
dE\,e^{iEt}\mkt{E}S(E)\pbk{E}{\phi};\;t\geq 0.\label{bgtls}
\end{equation}

These equations ((\ref{cfbexpmp}), (\ref{lsexp}) through
(\ref{bgtls})) are understood as a functional equation over all
$\mup{\psi}\in\pdo{\Phi}$.  This means that these expansions of
$\pup{\phi}(t)\in\mdo{\Phi}\subset\tup{\pdo{\Phi}}$ can be used to
obtain \bk{\mup{\psi}}{\pup{\phi}(t)} (whose modulus square is the
probability to find the time evolved state by a detector that detects
the observable \kb{\psi^-}{\psi^-}) for any $\psi^-\in\pdo{\Phi}$, but
{\em not} to calculate \bk{\psi}{\phi} for a $\psi\in\mdo{\Phi}$.
This is not a problem because the expressions
\norm{\bk{\psi}{\phi^+(t)}}, $\psi\in\Phi_-$ and $\phi^+\in\Phi_-$,
have no physical meaning, since they would represent the probability
for finding an in-state $\psi$ in an in-state $\phi^+(t)$ at $t \geq
0$ which is not measurable in a scattering experiment.  The semigroup
time evolution operator $\tup{\pdo{U}}(t)$ can also be applied to the
basis vector expansion (\ref{fbexpmp}) with \pup{\phi} understood as
functional $\pup{\phi}\in\tup{\mdo{\Phi}}$ over the
$\mup{\psi}\in\pdo{\Phi}$ because (\ref{fbexpmp}) is a functional
equation over \pdo{\Phi} since the \mbk{E}{\psi} are according to
(\ref{outspace}) elements of
${\cal S}\cap\pdo{\bohmHS}^{2}|_{\pup{{\Bbb R}}}$.  However the
semigroup operator $\tup{\pdo{U}}(t)$ cannot be applied to the
representation (\ref{fbexppb}) for
$\pup{\phi}\in\mdo{\Phi}\subset\tup{\pdo{\Phi}}$, because it is
{\em not} a functional equation over \pdo{\Phi} since the
\pmbk{E}{\psi}, with $\mup{\psi}\in\pdo{\Phi}$, are not necessarily
elements of ${\cal S}\cap\pdo{\bohmHS}^{2}|_{\pup{{\Bbb R}}}$.

Of the two exact but different basis vector expansions (\ref{fbexppb})
and (\ref{cfbexpmp}) for the same $\phi^+\in\Phi_-$, (\ref{fbexppb})
is the standard expansion and has a correspondence in the Hilbert
space (spectral resolution of operators with a continuous spectrum).
The expansion (\ref{cfbexpmp}) is new and shows that the
quasi-stationary states \mkt{z_{R_i}} can serve as basis vectors in
very much the same manner as the stationary states \dkt{E_n} in the
standard case.  But in addition to the resonance states the new basis
vector expansion (\ref{cfbexpmp}) (for any $N\neq 0$, e.g. $N=1$ or $N=2$ 
as in (\ref{tevolls})) also contains an integral over the
negative real axis from $E=-\infty_{II}$ to $0$ in the second sheet of
the energy surface of the $S$-matrix.  This integral depends on the 
preparation of the state and may be
infinitesimally small, but cannot be zero.  Its time dependence is 
non-exponential and could cause observable deviations from the exponential 
law for the transition rate of the prepared state $\phi^+$.  It may also have 
some other small but observable consequences.

The result (\ref{tevolls}) means that the semigroup time evolution of
a superposition of two (or more) Gamow states does not regenerate one
Gamow state from the background $\pbg(t)$ or from the other Gamow
vector.  In particular, if the state \pup{\phi} can be prepared such
that at some time $t_0\geq 0$ the background term $\pbg(t)$ is
practically zero, then it will remain practically zero for all
$t>t_0$, and the two Gamow states will evolve separately according to
separate exponential laws without regenerating each other:
\begin{equation}
\pup{\phi}(t)\approx e^{-i(E_{L}-i\Gamma_{L}/2)t}\pgl b_{L} +
e^{-i(E_{S}-i\Gamma_{S}/2)t}\pgs b_{S};\;\ t\geq 0.\label{appls}
\end{equation}
Approximations like (\ref{appls}) have been used for the time
evolution of one- and two-resonance systems (like the $K_L$-$K_S$
system with $\phi^+(t)$ representing the $K^0$ state) in theories with
``effective Hamiltonians'' given by a $2\times 2$ complex
diagonalizable matrix.

The full expansion (\ref{cfbexpmp}) (again neglecting bound states)
leads to a matrix representation of the self-adjoint semi-bounded
Hamiltonian $H$ in the following form:
\pagebreak
\begin{eqnarray*}
\left( \begin{array}{c} \bk{H\psi^-}{z_{R_1}^-} \\
\bk{H\psi^-}{z_{R_2}^-} \\ \vdots \\ \bk{H\psi^-}{z_{R_N}^-} \\
\bk{H\psi^-}{E^-} \end{array} \right)& =& \left( \begin{array}{c}
\mate{\psi^-}{\tup{H}}{z_{R_1}^-}\\
\mate{\psi^-}{\tup{H}}{z_{R_2}^-}\\ \vdots \\
\mate{\psi^-}{\tup{H}}{z_{R_N}^-}\\ \mate{\psi^-}{\tup{H}}{E^-}
\end{array} \right)\\
&=& \left( \begin{array}{cccccc} z_{R_1} & & & & 0
\\ & z_{R_2} & & & \vdots \\ & & \ddots & & \vdots \\ & & & z_{R_N} &
0 \\ 0 & \cdots & \cdots & 0 & (E) \end{array} \right) \left(
\begin{array}{c} \bk{\psi^-}{z_{R_1}^-} \\ \bk{\psi^-}{z_{R_2}^-} \\
\vdots \\ \bk{\psi^-}{z_{R_N}^-} \\ \bk{\psi^-}{E^-} \end{array}
\right)
\end{eqnarray*}
\begin{equation}
\;\;\;\;\;\;\psi^-\in\Phi_+\subset\Phi,\;\; -\infty_{II}<E\leq
0,\label{mateHsp}
\end{equation}
where the lowest row represents the diagonal continuously infinite
real energy matrix:
\begin{equation}
\left( \bk{H\psi^-}{E^-} \right) = \left( \mate{\psi^-}{\tup{H}}{E^-}
\right) = \left( E \right) \left( \bk{\psi^-}{E^-} \right);\
\psi^-\in\Phi,\ -\infty_{II}<E\leq 0.\label{cirem}
\end{equation}
Since the basis vector expansion (\ref{cfbexpmp}) is an exact
representation of $\phi^+\in\Phi_-$, the matrix representation
(\ref{mateHsp}) also is an exact representation of the self-adjoint
Hamiltonian.  In the phenomenological descriptions by complex
effective Hamiltonians, one uses a truncation of (\ref{mateHsp}) which
corresponds to omitting the background integral, i.e., omitting the
whole continuously infinite diagonal matrix $(E)$ (and sometimes even
some of the $z_{R_i}$).

The extra term \pbg\ in (\ref{lsexp}) and (\ref{tevolls}) is not taken
into consideration in any of the finite dimensional effective theories
of complex Hamiltonians, and in particular not in the Lee-Oehme-Yang
theory of the neutral Kaon system.  This term, which comes from the
integral along the negative real axis in the second sheet of the
$S$-matrix, can be shown to be also decaying, i.e.,
$\norm{\bk{\psi^-}{\pbg(t)}}\ \rightarrow\ 0$ for $t\ \rightarrow\
\infty$ for every $\psi^-\in\Phi_+$, but it decays more slowly than
the exponential.  The standard effective theories, like the enormously
successful Lee-Oehme-Yang theory of the neutral K-system, emerge as
subtheories of the exact complex basis vector expansion
(\ref{cfbexpmp}) or (\ref{lsexp}) in the $N$-dimensional space
${\cal M}_N=\left\{ \phi | \phi=\sum_{i=1}^N \mkt{z_{R_i}}c_i
\right\}$ spanned by the Gamow vectors \mkt{z_{R_i}}.  However
${\cal M}_N$ is a subspace of $\Phix_+$ and lies {\em outside} the
standard HS.  These effective theories are usually legitimized by the
Wigner-Weisskopf approximation.  In our irreversible quantum theory
the expression (\ref{tevolls}) is exact and can be used to justify the
effective theory (\ref{appls}) as a subtheory in
${\cal M}_N\subset\Phix_+$ which remains invariant under time
translations in the forward direction, (\ref{tevolls}).  The
expression (\ref{tevolls}) shows that the ``deviation from the
exponential decay law'' does not arise for the resonance state but are
the properties of the background terms (background phase shifts).

The emergence of such an enormously successful phenomenological
description as the Lee-Oehme-Yang theory as a subtheory is an
empirical validation of the complex basis vector expansion
(\ref{cfbexpmp}).

\subsection{The Golden Rule from Fundamental Probabilities
\label{Thegold}}

\footnotesize
\begin{quote}
``[T]he physical system leaves the [initial decaying] state \kt{\phi_i}
irreversibly.''

---Cohen-Tannoudji, et al.\footnote{p. 1345 of \cite{C-T}.}
\end{quote}
\normalsize

The probabilities ${\cal P}(t)=\mathrm{Tr}(\Lambda W)$ are the most
fundamental quantities in quantum physics.  They describe the probability
to register or measure the observable $\Lambda$ in the state $W$ and
represent the directly measured experimental quantities; 
${\cal P}(t)$ is measured by the number of counts of a registration 
apparatus (detector) and its derivative, 
$\dot{{\cal P}}=\frac{d}{dt}{\cal P}(t)$ is observed as the 
normalized counting rate of the detector.  We want to apply these
probabilities to the case where $\Lambda$ is a projection operator (or
more generally a positive definite operator) on a subspace of non-interacting
decay products registered by a detector and $W$ is a 
quasistable state $W^D(t)$.  In this case ${\cal P}(t)$
is the decay probability and $\dot{{\cal P}}(t)$ is the decay rate
for the decaying state $W^D$ into the decay products $\Lambda$ registered
by the detector.  Thus $\Lambda$ is experimentally defined by the detector
and $W^D(0)$ by the preparation apparatus or preparation process of
the quasistable state (e.g. a resonance produced in a scattering experiment
or a metastable product of some ancient creation process
which is present at some initial time 
$t=0$).

Traditionally, $W$ is an
ensemble defined by a macroscopic preparation apparatus (single microsystems 
are not defined) and $\Lambda$ and $W$ are mathematically
represented by operators in the Hilbert space.  However, as
we mentioned in Section 3, the decay probabilities within the HS
formulation have severe problems, justifying a fresh approach.
  Our approach is to extend the basic formula 
${\cal P}(t)=\mathrm{Tr}(\Lambda W)$
to apply to the decaying state $W^D$ mathematically represented by the 
Gamow vector of Section \ref{Gamowv},
$W^D=\kb{\psi^G}{\psi^G}$.  We give the following interpretation to this
basic formula applied to $W^D$: a single microphysical decaying system
described by $W^D$ has been produced by a macroscopic apparatus
and a quantum scattering process, at a time $t=0$.  Each count of the
detector in a decay experiment is the result of the decay of this single 
microsystem that has lived
for a time $t_a$---the time that 
it took the decaying system 
to travel from the scattering center to the decay vertex.
The integral 
\[
\int_{t_a-\frac{1}{2}\Delta t}^{t_a+\frac{1}{2}\Delta t}dt\:
\dot{{\cal P}}(t)\approx\dot{{\cal P}}(t_a)\Delta t, 
\]
is proportional to the number of microsystems that have decayed in the
time interval $\Delta t$, i.e. $\dot{{\cal P}}(t)$ 
is the normalized counting rate of the detector, $\dot{N}(t)/N(\infty)$.  
The lifetime $\tau_D$ (mean life) is the average of the
$t_a$'s:
\[
\sum_a t_a\: \Delta t\:\dot{{\cal P}}(t_a)\approx\int dt\:t\:
\dot{{\cal P}}(t)=\tau_D.
\]
This is the way $\dot{{\cal P}}(t)$, $\tau_D$, 
etc. are obtained (defined) 
experimentally.  The quantity ${\cal P}(t)$ in the 
basic formula is thus experimentally the normalized number of counts
$N(t)/N(\infty)$ (with $N$ very large).  Therefore the normalization condition
of the $W^D$ is such that
\begin{equation}
{\cal P}(\infty)=1;
\end{equation}
the probability of finding the state decayed is certainty.  In addition 
${\cal P}(t)$ is subject to an initial boundary condition
\begin{equation}
{\cal P}(0)=0;
\end{equation}
the probability of finding one of the many ($N$) decay products already at 
$t=0$ is zero.  The reason for this is that $t=0$ is the time at which the 
decaying state has been prepared and one starts counting the decay products
(one assumes there are no other incoming decay products in the detector area).

In the mathematical theory
of quantum mechanics, the quantity ${\cal P}(t)$ is calculated from
the operators $\Lambda$ and $W^D$ by the basic formula (\ref{latergold}) given
by
\begin{equation}\label{decprob}
{\cal P}(t)=\mathrm{Tr}(\Lambda W^D(t))=\mathrm{Tr}(\Lambda(t)W^D)=
\mate{\psi^D(t)}{\Lambda}{\psi^D(t)},\ \mbox{for}\ t\geq 0\ \mbox{only}.
\end{equation}
The formula for the decay rate $\dot{{\cal P}}(t)$ should be obtained
as the derivative of the right hand side of (\ref{decprob}).

However, in the standard textbooks of quantum mechanics
 one does not calculate the transition probabilities (\ref{decprob})
\footnote{One cannot calculate it in the HS formulation because of 
(\ref{impossdec}), \cite{Heger}.}.
 Instead one gives a Golden Rule~\cite{Fermi} for the initial decay rate
 $\dot{{\cal P}}(t=0)$, and one justifies this Golden Rule
with some ingenious heuristic 
arguments that were originated by Dirac~\cite{Goldberg}.  In place of
 an initial (t=0) decaying state $W^D=\kb{\psi^D}{\psi^D}$ 
which evolves by the exact Hamiltonian $\psi^D(t)=e^{-iHt}\psi^D$, one usually 
chooses an eigenstate of the interaction-free 
Hamiltonian, $\psi^D  \rightarrow  f^D$, where
\begin{equation}\label{i-free}
H_0 f^D = E_D f^D\ \mbox{with}\ H_0= H-V.
\end{equation}
  In place of the projection operators 
for decay products, one uses the ``improper'' states, 
$\Lambda \rightarrow \kb{E_b, b}{E_b, b}$ which have dimension of
$(\mbox{energy})^{-1}$.  The Golden Rule is then given as 
the decay rate energy-density
\begin{subequations}\label{texbok}
\begin{equation}\label{texboka}
\dot{{\cal P}}^{E_b}_{b\:\:\:\:D}(0) = \frac{2\pi}{\hbar}
\norm{\mate{b, E_b}{V}{f^D}}\delta(E_D - E_b),
\end{equation}
and 
the initial decay rate is obtained by integration and/or summation over
all final quantum numbers $E_b$ and $b$, as:
\begin{equation}\label{BAeGR}
\dot{{\cal P}}_D(0)= \frac{2\pi}{\hbar} \int \: dE_b \sum_b
\norm{\mate{b, E_b}{V}{f^D}}\delta(E_b - E_D).
\end{equation}
Alternatively, 
 one chooses for the initial state also an improper ``energy eigenstate'' of
the free Hamiltonian, $H_0\kt{E_D,d}=E_D\kt{E_D,d}$, or of the exact 
Hamiltonian, $H\pkt{E_D,d}=E_D\pkt{E_D, d}$.  Then, for example, 
one can write the exact but highly singular Golden Rule as
\begin{eqnarray}\label{inttexboka}
\dot{{\cal P}}^{E_b E_D}_{b\:\:\:\:D}(0)& =& \frac{2\pi}{\hbar}
\norm{T_{bD}}\delta(E_D - E_b), \mbox{where}\\
T_{bD}&=&\br{b,E_b}V\pkt{E_D, d}\approx\mate{b,E_b}{V}{E_D, d}.\nonumber
\end{eqnarray}
\end{subequations}
The initial transition probability rate $\dot{{\cal P}}(0)$ is then 
obtained by integration/summation over the final quantum numbers
 $E_b$ and $b$ and averaging over the initial quantum numbers $E_D$ and $d$.

The expression (\ref{BAeGR}) (and its variants)
 is one of the most important and most widely used formulas in 
quantum physics.  It expresses the decay probability per unit time  of the 
state $\kb{f^D}{f^D}$ at $t=0$ 
(the time when the decaying state has been created and the registration of 
the decay products begins) into the 
non-interacting decay products described by the projection operator given by
$\Lambda$:
\begin{equation}
\Lambda = \sum_{\stackrel{\mathrm{all}\ b}{b\neq b^D}}\:\int_0^\infty
dE\:\kb{E,b}{E,b},\;\ \mbox{where}\ \ H_0\kt{E,b}=E\kt{E,b}
\end{equation}

The initial decay rate 
$\dot{{\cal P}}(t=0)=\frac{d}{dt}{\cal P}(t)|_{t\rightarrow 0^+}$ of
(\ref{BAeGR}) should be the time derivative of the probability
(\ref{decprob}), at least approximately, or it should be the time 
derivative of a probability 
${\cal P}(t)=\mathrm{Tr}(\Lambda(t)\kb{f^D}{f^D})$.
This suggests the following program: Find a decaying state 
$W^D(t)=\kb{\psi^D(t)}{\psi^D(t)}$ that evolves in time according to the
exact Hamiltonian $H=H_0+V\geq 0$, 
\begin{equation}
W^D(t)=e^{-iHt}W^D(0)e^{iHt},
\end{equation}
calculate the decay probability ${\cal P}(t)$ using the basic formula
(\ref{decprob}), and then take the time derivative of the probability, 
$\dot{{\cal P}}(t)$.  This decay rate at $t=0$, $\dot{{\cal P}}(0)$,
should somehow resemble Dirac's Golden Rule (\ref{BAeGR}), at least in some
approximation $\psi^D\rightarrow f^D$.

From the results discussed in Sect. \ref{Conseq}, 
in particular from the result that
${\cal P}(t)\equiv 0$ for every $\psi^D\in\bohmHS$~\cite{Heger}, it is clear
that such a program cannot be implemented by a mathematical theory in the 
Hilbert space.  The futility of such attempts was the actual reason for the
introduction of the Gamow vectors~\cite{Bohm}.  We shall now describe how
the Gamow vectors and irreversible quantum mechanics in the RHS lead to
a decay rate formula that reduces to Dirac's Golden Rule in the
Born approximation.

For the detector $\Lambda$, one takes a positive operator
\begin{equation}\label{Lambda}
\Lambda = \sum_{\stackrel{\mathrm{all}\ b}{b\neq b^D}}\:\int_0^\infty
dE\:\lambda_b(E)\kb{E,b}{E,b},
\end{equation}
where $H_0\kt{E,b}=E\kt{E,b}$ and where $\lambda_b(E)$ is a smooth and 
rapidly decreasing function of $E$ 
(and of all the other (continuous) quantum numbers in $b$) that describes the
detector efficiency.  The summation in $b$ extends over those 
quantum numbers (and momentum directions) of the decay products which are 
registered by the detector but the summation 
does not include the quantum numbers
$b^D$ of the decaying state.

For the decaying state 
$W^D(t)=\kb{\psi^D(t)}{\psi^D(t)}$ one takes the pure Gamow state $\psi^G$
of Sect. \ref{Gamowv}:
\begin{equation}\label{DtoG}
\psi^D(t)=\frac{1}{f}\psi^G(t)\in\Phix_+,
\end{equation}
where the ``normalization'' factor $f$ is chosen such that 
${\cal P}(\infty)=1$ when $\Lambda$ describes {\em all} decay products 
This state has the time evolution (\ref{decayst}),
which in operator form is:
\begin{subequations}\label{GRevol}
\begin{equation}\label{GRevola}
W^G(t)=\kb{\psi^G(t)}{\psi^G(t)}=e^{-i\tup{H}t}\kb{\psi^G}{\psi^G}e^{iHt}
=e^{-\Gamma t}\kb{\psi^G}{\psi^G},\;\mbox{for}\ t\geq 0\ \mbox{only}.
\end{equation}
This is inserted for $W^D(t)$ in (\ref{decprob}) where also $t\geq 0$.
The equality (\ref{GRevola}) is understood as a functional equation in the 
space $\Phi_+$, as in 
(\ref{decayst}) and (\ref{decaystcc}), i.e. as
\begin{equation}
\mate{\psi_1^-}{W^G(t)}{\psi^-_2}=e^{-\Gamma t}\bk{\psi_1^-}{\psi^G}
\bk{\psi^G}{\psi_2^-}\;\ 
\mbox{for}\ \psi_1^-,\psi_2^-\in\Phi_+.
\end{equation}
\end{subequations}
This means (\ref{GRevola}) is only valid in matrix elements with the 
vectors $\psi^-\in\Phi_+$, which represent the out-states that are 
ultimately registered by the detectors.  This presents no limitations, since
only those matrix elements are experimentally accessible\footnote
{The matrix element $\mate{\phi^+}{W^G}{\phi^+}$ does not make any sense
mathematically or physically.}.

The decay probability (\ref{decprob}), with (\ref{Lambda}) and (\ref{DtoG}),
can now be derived in a lengthy calculation~\cite{Bohm} using (\ref{GRevol}).
This derivation makes use of the Lippmann-Schwinger equation 
in one of its standard (singular) forms:
\begin{equation}\label{L-S}
\mkt{E,b}=\kt{E,b}+\frac{1}{E-H-i\epsilon}V\kt{E,b}
\end{equation}
where $\tup{H}\mkt{E,b}=E\mkt{E,b}$ and $\tup{H}_0\kt{E,b}=E\kt{E,b}$.
The derivation also makes use of the relation
\begin{equation}\label{functeq}
\mate{{}^-E}{f(\tup{H})}{\psi^G}=f(z_R)\bk{{}^-E}{\psi^G},
\end{equation}
which follows from (\ref{actofH}) and can be proven as a functional 
equation over the
$\mbk{\psi}{E}\in({\cal S}\cap\bohmHS^2_-)_{{\Bbb R}^+}$.  From
(\ref{L-S}) and (\ref{functeq}) one obtains for sufficiently good interaction
Hamiltonians $V$ (such that $\mate{E}{V}{\psi^G}\in
({\cal S}\cap\bohmHS^2_-)^\times$):
\begin{equation}\label{toosing}
\bk{\psi^G}{{E,b}^-}=\bk{\psi^G}{E,b}+\mate{\psi^G}{V}{E,b}
\frac{1}{E-(E_R+i\Gamma/2)-i\epsilon},
\end{equation}
This is again understood as a functional equation over the 
$\mbk{\psi}{E}\in({\cal S}\cap\bohmHS^2_-)_{{\Bbb R}^+}$.
With these equations, and under the assumption that the mathematically 
singular expressions above can be rigorously justified\footnote{
In contrast to the statements 1.\ and 2.\ (eq.\ref{impossdec}) in 
Sect.\ \ref{Conseq}), whose proofs use only well defined mathematics of the 
HS, the statement (\ref{GRinter}) cannot be formulated as a mathematical 
theorem because its derivation requires such singular expressions as the 
Lippmann-Schwinger equation.  While this equation is exact and 
well-accepted by the physics
community, it has as yet not been given a mathematically rigorous 
foundation in either the HS or RHS.  In this respect (\ref{GRinter}) and 
(\ref{eGR}) are also different from the statements in the preceeding sections,
like the semigroup time evolutions (\ref{inU}) and (\ref{outU}) in 
$\Phi_\mp\subset\bohmHS\subset\Phix_\mp$, the exponential law (\ref{decst}) and 
(\ref{GRevol}), the Nuclear Spectral Theorem (\ref{fbexpp}) and the
``complex spectral theorem'' (\ref{cfbexpmp}) and (\ref{mateHsp}), which 
are well-founded in the mathematics of the RHS.  On the other hand, 
(\ref{eGR}) reproduces in a reasonable approximation the well-proven Golden
Rule, providing heuristic support.}, one calculates the 
following result:
\begin{equation}\label{GRinter}
{\cal P}(t)=1-e^{-\Gamma t}\int_0^\infty\:dE\:\sum_{b\neq b^D}
\lambda_b(E)\norm{\mate{E,b}{V}{\psi^G}}
\frac{1}{(E-E_R)^2 + (\Gamma / 2)^2};\;t\geq 0.
\end{equation}
This is the probability for the transition of the decaying state $W^D$ into all
mixtures of decay products with the property $\Lambda$.  Usually one omits
the detector efficiency $\lambda_b(E)$ in formulas like (\ref{GRinter}) above
and (\ref{eGR})$\ldots$(\ref{bound}) and (\ref{Bornres}) below.  This means one
gives these formulas for an ideal 
detector with $\lambda_b(E)=1$ for all $b$ which are quantum numbers of decay 
products and with $\lambda_b(E)=0$ otherwise.  The detector efficiency 
$\lambda_b(E)$ is then used in the analysis of the experimental
data to correct the observed events for detector efficiency.  The reported
counting rate has usually been corrected for this detection efficiency.  
The factor $\lambda_b(E)$ is then
omitted in the theoretical formulas like (\ref{GRinter}), 
but this factor is always present and can be
used in the mathematical calculations to make
expressions like (\ref{toosing}) less singular.

Taking the time derivative of (\ref{GRinter}) (setting
$\lambda_b(E)=1$), we obtain for the decay rate
\begin{equation}\label{eGR}
\dot{{\cal P}}(t)=e^{-\Gamma t}2\pi\int_0^\infty\:dE\:\sum_{b\neq b^D}
\norm{\mate{E,b}{V}{\psi^G}}
\frac{\Gamma/2\pi}{(E-E_R)^2 + (\Gamma / 2)^2};\;\;t\geq 0.
\end{equation}
Since $\lambda_b(E)$ has been omitted the decay rate in (\ref{eGR}) is to be 
compared to the experimental counting rate corrected for detection efficiency.
The formula (\ref{eGR}), and also (\ref{GRinter}),
 we call the exact Golden Rule.  

The initial decay rate is then obtained as
\begin{equation}\label{eGRinit}
\dot{{\cal P}}(0)=2\pi\int_0^\infty\:dE\:\sum_{b\neq b^D}
\norm{\mate{E,b}{V}{\psi^G}}
\frac{\Gamma/2\pi}{(E-E_R)^2 + (\Gamma / 2)^2}.
\end{equation}
Since the probability to find the decay product at time $t\leq 0$ needs
to be zero, ${\cal P}(0)=0$, we obtain from (\ref{GRinter}) for
$t=0$:
\begin{equation}\label{bound}
\frac{2\pi}{\Gamma}\int_0^\infty\:dE\:\sum_{b\neq b^D}
\norm{\mate{E,b}{V}{\psi^G}}
\frac{\Gamma/2\pi}{(E-E_R)^2 + (\Gamma / 2)^2}=1.
\end{equation}
The sums in (\ref{GRinter})$\ldots$(\ref{bound}) extend over all quantum 
numbers and all decay products.  Comparing (\ref{bound}) with 
(\ref{eGRinit}), we obtain
\begin{equation}\label{rate}
\dot{{\cal P}}(0)=\Gamma\;\;\;(=\Gamma/\hbar),
\end{equation}
and from the exponential time dependence in 
(\ref{GRinter}) or (\ref{eGR}) we obtain the lifetime of the decaying state:
\begin{equation}
\tau_G=\frac{1}{\Gamma}\;\;\;(=\hbar/\Gamma),
\end{equation}
The result (\ref{rate}) means that the imaginary part of the complex 
energy in (\ref{actofH}), 
which is also the imaginary part of the $S$-matrix pole position
and  the width of the 
Breit-Wigner energy distribution (\ref{BWd}) is equal to the
initial rate
$\dot{{\cal P}}(0)$ of the decay probability (\ref{decprob}).

The formulas (\ref{GRinter}) and the formulas (\ref{eGR}) and (\ref{eGRinit})
give the {\em total} decay probability and the {\em total} 
decay rate, respectively, because we took
for $\Lambda$ the projection operator on {\em all} decay products and summed
over {\em all} values of quantum numbers $b=\{b_1,b_2,\ldots,\eta\}$ of 
{\em all} the decay products.  The partial
decay rates are obtained if we sum in (\ref{eGR}) and (\ref{eGRinit}) only
over part of the quantum numbers $b=\{b_1,b_2,\ldots,\eta\}$, e.g. over
all $\hat{b}=\{b_1,b_2,\ldots\}$ but not over the quantum number $\eta$ that
 characterizes the decay channels:
\begin{equation}\label{eGRinitpart}
\dot{{\cal P}}_\eta(t)=2\pi e^{-\Gamma t}
\int_0^\infty\:dE\:\sum_{b_1,b_2,\ldots}
\norm{\mate{E,b_1,b_2,\ldots,\eta}{V}{\psi^G}}
\frac{\Gamma/2\pi}{(E-E_R)^2 + (\Gamma / 2)^2}.
\end{equation}
The value $\dot{{\cal P}}_\eta$ is the partial decay rate into decay 
products with the quantum number $\eta$ or into the decay channel 
characterized by $\eta$.  The total decay rate ((\ref{eGR}) or 
(\ref{eGRinit})) is then written as the sum over partial decay rates, with 
a corresponding relation for the ``partial widths'' 
$\Gamma_\eta\equiv\hbar\dot{{\cal P}}_\eta(0)$:
\begin{equation}\label{totwid}
\dot{{\cal P}}(t)=\sum_{\mbox{\tiny all}\ \eta}
\dot{{\cal P}}_\eta(t),\;\;\;\;\Gamma=\sum_{\mbox{\tiny all}\ \eta}
\Gamma_\eta.
\end{equation}
In analogy to (\ref{totwid}), with (\ref{eGRinitpart}) one can write
for the decay probability (\ref{GRinter}) (setting $\lambda_b(E)=1$)
\begin{multline}
{\cal P}(t)=1-\sum_\eta \frac{2\pi}{\Gamma}
e^{-\Gamma t}\int_0^\infty dE\sum_{b_1,\ldots\neq b^D}
\norm{\mate{E,b_1,\ldots\eta}{V}{\psi^G}}\\
\times\frac{\Gamma / 2 \pi}{(E-E_R)^2 + (\Gamma / 2)^2},\;\;\;\;t\geq 0.
\end{multline}
The derivation of the above formulas (\ref{GRinter}), (\ref{eGR}), 
(\ref{eGRinit}) and (\ref{eGRinitpart}) did not make use of any approximations.
 We therefore call these formulas exact Golden Rules.
Equation (\ref{GRinter}) is the exponential decay law with the 
directionality of time~\cite{WW}
\begin{equation}
{\cal P}(t)=1 - e^{-\Gamma t},\;\;\;t\geq 0.
\end{equation}
From the quantum mechanical arrow of time (Section 4), via the semigroup 
(\ref{outU}) and the evolution of the Gamow states (\ref{decayst}), we
derived the time-asymmetry for the decay process: {\em If we start at $t=0$
from a ``decaying state'' $\kb{\psi^G}{\psi^G}$, then the probability of 
finding the decay products $\Lambda$ increases exponentially with time from 
${\cal P}(0)=0$ to ${\cal P}(t\rightarrow\infty)=1$.}

In addition to these results concerning principles, we can also obtain a
rule for the calculation of the initial decay rate or width in quantum
theory.  The exact Golden Rules are not
of much practical use in calculations since these formula contain the 
unknown $\psi^G$ and the natural line width under the integral.  
The state vector 
$\psi^G$ is unknown and the values ($E_R,\Gamma$) are unknown until
the eigenvalue equation
\begin{equation}
\tup{H}\psi^G=\left(E_R - i\frac{\Gamma}{2}\right)\psi^G;\;\;\psi^G\in\Phix_+
\end{equation}
has been solved.  Therefore, to obtain a calculational tool from the above 
relations (\ref{eGR}), (\ref{eGRinit}) and (\ref{eGRinitpart}),
one uses the Born approximation on the right hand side: one expands the 
exact $\psi^G$ in a perturbation series 
(for the interaction Hamiltonian $V$) in terms of the non-interacting
 ``decaying state vector'' $f^D$ defined by (\ref{i-free}), since the 
solutions of (\ref{i-free}) are usually known.

The Born approximation is given by
\begin{subequations}\label{Bornapp}
\begin{eqnarray}
\mate{b,E}{V}{\psi^D}&\approx&\mate{b,E}{V}{f^D}\\
\frac{\Gamma}{2E_R}&\rightarrow&0\\
E_R&\approx &E_D
\end{eqnarray}
The Breit-Wigner energy distribution (natural line width) has the property:
\begin{equation}\label{limitBW}
\lim_{\frac{\Gamma}{2E_R} \rightarrow 0} 
\frac{\Gamma/2\pi}{(E-E_R)^2 + (\Gamma / 2)^2}=\delta(E-E_R).
\end{equation}
\end{subequations}
Using (\ref{Bornapp}), one obtains from (\ref{eGRinit})
the initial decay rate in this Born approximation:
\begin{equation}\label{Bornres}
\dot{{\cal P}}(0)= \frac{2\pi}{\hbar} \int \: dE \sum_b
\norm{\mate{b, E}{V}{f^D}}\delta(E - E_D),
\end{equation}
and similar expressions for the other formulas.  
This is the standard Golden Rule (\ref{BAeGR}) again, only here it has been 
{\em derived} from the transition probability (\ref{decprob}) and from 
the time evolution 
of the Gamow vectors (\ref{GRevol}) as the Born
approximation of an exact Golden Rule.

The Gamow vector $\psi^G$ provides the link that was 
missing from the HS
formulation to connect the all-important empirical rules for the rates to the
fundamental theoretical relations for the probabilities.

\section{Summary}

Dirac's bra and ket formalism can be given a mathematical justification
by the
rigged Hilbert space (RHS).  The RHS does not
only contain Dirac kets (``scattering states'') but also Gamow
vectors, which are eigenkets \kt{E_R-i\Gamma/2} of self-adjoint
Hamiltonians $H$ with complex eigenvalue ($E_R - i\Gamma / 2$)
representing decaying states or resonances with Breit-Wigner energy
distribution of energy $E_R$ and width $\Gamma$.  The value $\Gamma$
is also the initial rate of the decay probability ${\cal P}(t)$,
$\Gamma = \hbar \frac{d{\cal P}}{dt}|_{t=0}$, for which an exact
Golden Rule could be derived.  In the Born approximation, this exact 
Golden Rule  becomes
Dirac's Golden Rule.  
These features are central
requirements for a theory of resonances and decay, yet the standard
quantum theory in Hilbert space (HS) can not produce them 
since the transition probabilities from a decaying state $W^D(t)$ into the 
observed decay products $\Lambda$, ${\cal P}(t)=\mathrm{Tr}
(\Lambda W^D(t))$, can be proven to be 
identically zero if they are zero at any time interval,
 e.g.~before the decaying state has been prepared.  Therefore, resonances 
in the HS formulation of quantum mechanics cannot be described by a 
state vector $\psi^D\in\bohmHS$ and had to be 
denied the status of autonomous microphysical
entities in HS.  This is contrary to the phenomenological observation that 
stability and the value of the lifetime are
not criteria for elementarity and that 
quasistable states and stable states should be described on the 
same footing---as it is done in $S$-matrix theory.

The semigroup time evolution of the Gamow states sprung from the new
mathematical language of the RHS and led to their exact exponential
decay with lifetime $\tau=\hbar/\Gamma$
\[e^{-iH^{\times}t}\kt{E_R - i\Gamma/2}=
e^{-iE_Rt}e^{-(\frac{\Gamma}{2}) t}\kt{E_R - i
\Gamma/2},\ \mbox{for}\ t\geq 0\ \mbox{only}.
\]
The semigroup $e^{-i H^{\times} t}$, $t \geq 0$, expresses intrinsic
irreversibility on the microphysical level.  This irreversibility
is not restricted to the time evolution of Gamow states.  It can be 
formulated in terms of time-asymmetric boundary
conditions for states and observables and is related to a general
arrow of time which can be expressed by the truism: A state
needs to be prepared first before an observable can be registered in
it.  This can also be expressed as a condition of time ordering $t_a>t_0$ in 
the probabilities ${\cal P}(t)=\mathrm{Tr}
(\Lambda(t_a) W(t_0))$, where $t=t_0$ is the time at which state $W$ has 
been prepared and the registration of the observable $\Lambda$ can begin 
(or the initial time of the universe $W(t_0)=\rho_i$ if the quantum 
mechanics of measured systems is extended to the quantum theory of
cosmology).
This intrinsic irreversibility seems to have a different origin than the
standard extrinsic irreversibility of open quantum systems which
results from the effect of an external reservoir or measurement
apparatus.  The latter has also a semigroup evolution which however is
generated by a reservoir-dependent Liouvillian and not by the
Hamiltonian of the quantum system.  This intrinsic irreversibility
also does not follow from the collapse axiom of a pure or less 
mixed state into a more
mixed state.  However, the change of state due to measurement scattering
of a microsystem on a macrosystem possesses, like every scattering process,
an arrow of time which has the same origin as the semigroup.

Since our time-asymmetry comes from the boundary conditions and 
the algebras of observables and the dynamical laws are
time-symmetric, microphysical irreversibility is compatible with a
time reversal transformation (or CPT) if one chooses for $T$ a
representation that doubles the space of states and observables.
Such representations of $T$
exist, but the interpretation of a time-reversal doubled world is not clear.

The Gamow kets describing Breit-Wigner resonances have been derived
from first order poles of the analytically continued $S$-matrix.  This
can be generalized to quasi-stationary systems associated with higher
order poles of the $S$-matrix.  An $S$-matrix pole of $r$-th order at
$z_R=E_R-i\Gamma/2$ leads to $r$ generalized eigenvectors of order
$k=0,1,\ldots,r-1$.  The Gamow vector of order $k$, $\kt{z_R}^{(k)}$,
is the $k$-th derivative of the ordinary Gamow vector and, except for
a normalization factor, also a Jordan vector of degree $(k+1)$ with
generalized eigenvalue $z_R = E_R - i\Gamma/2$.

The Gamow vectors appear in the complex basis vector expansion, which
is, for state vectors, 
 an alternative to the Dirac basis vector expansion and contains
Gamow kets instead of Dirac kets as basis
vectors.  The complex basis vector expansion is particularly useful
for problems that involve resonances and decay.  For a system with a
finite number of resonance states (such as the neutral Kaon system),
it leads to the ``effective'' theories with complex Hamiltonian
matrices (e.g. Lee-Oehme-Yang theory of the $K_L$-$K_S$ system) by
truncating the complex basis vector 
expansion to the subspace spanned by the Gamow vectors,
the resonance subspace.  In addition to the Gamow kets the complex
basis vector expansion of a prepared state vector 
also contains other terms called background.  This background term
varies with the preparation process of the state and its properties are
not connected with the decaying Gamow state (or states, if more than one
resonance is involved).  In particular, the time evolution of this 
preparation-dependent background term is non-exponential.

All that has been achieved by the RHS is a mathematical theory that allows to 
separate out the exponentially decaying (and also the exponentially growing 
which have not been discussed here) states and to derive their Golden Rule 
from the fundamental probabilities of quantum mechanics.  In the 
process of doing this, the time-asymmetry of quantum physics became apparent.

\noindent\Large{\textbf{Acknowledgment}}

\normalsize

\vspace{.5cm}

We would like to express our gratitude to the many friends, colleagues and 
teachers who helped us shape this paper into its present form.  
J.-P.~Antoine, M.~Gadella and O.~Melsheimer read various stages of the 
manuscript, corrected mistakes and suggested changes.  I.~Antoniou, 
Y.~Ne'eman and S.~Wickramasekara read different parts and made 
valuable suggestions.  C.~P\"untmann and M.~Loewe contributed to 
problems addressed by this article.  Discussions with J.~B.~Hartle,
G.~Ludwig and J.~A.~Wheeler about the arrow of time led to the writing 
(and rewriting) of the preface.  H.~Atmanspacher's historical expertise led
to the inclusion of the literature quotations.  Various topics addressed in 
this paper were discussed with the participants of the workshop on
Recent Problems of Quantum Mechanics and Cosmology of the Fondation
Peyresq Foyer d'Humanisme, Peyresq, St.~Andr\'e-les-Alpes, France.  
Their hospitality, and the 
financial support of NATO, is gratefully acknowledged.  We are appreciative
of the Welch Foundation for their financial support.

\end{document}